\documentclass[aps,pre,epsfig,floats,twocolumn,superscriptaddress,amssymb,amsmath,floatfix,showpacs]{revtex4}
\usepackage{graphicx}
\usepackage{bm}
\usepackage{amssymb}
\usepackage{color}
\usepackage{amscd}

\begin{document}
\title{Casimir rack and pinion: Mechanical rectification of periodic multi-harmonic signals}

\author{Amin Dehyadegari}
\affiliation{Department of Physics, University of Tehran, P. O. Box 14395-547, Tehran, Iran}

\author{MirFaez Miri}
\email{mirfaez_miri@ut.ac.ir}
\affiliation{Department of Physics, University of Tehran, P. O. Box 14395-547, Tehran, Iran}
\affiliation{School of Physics, Institute for Research in Fundamental Sciences, (IPM) Tehran 19395-5531, Iran }

\author{Zahra Etesami}
\affiliation{Department of Physics, University of Tehran, P. O. Box 14395-547, Tehran, Iran}

\begin{abstract}
We study noncontact rack and pinion composed of a corrugated plate and a corrugated cylinder
intermeshed via the lateral Casimir force.
We assume that the rack position versus time is a periodic multi-harmonic signal.
We show that in a large domain of parameter space and at room temperature, the device acts as a mechanical rectifier:
The pinion rotates with a nonzero average velocity and lifts up an external load. The thermal noise may even facilitate the device operation.
\end{abstract}

\pacs{07.10.Cm,85.85.+j,42.50.Lc,46.55.+d}

\maketitle
\section{Introduction}

In the realm of electronic circuits, the key role of {linear} resistors, capacitors, and inductances is clear. But
the electronic circuits have grown to maturity since the introduction of active and {\it nonlinear} elements.
Indeed {\it rectifiers}, voltage limiters, bistable multivibrators, frequency mixers, etc.,  are not realizable
without nonlinear components such as {\it diodes} and transistors\ \cite{Horowitz,Hess}. Microelectromechanical systems (MEMS) technology has its roots in microelectronics technology, thus naturally much attention has been paid to the nonlinear behavior of MEMS for use in signal processing, actuation, and sensing applications\ \cite{bookmem1,turner}.

With the ongoing trend of miniaturizing devices to the submicron scale, the gap between surfaces
may about $100 \;{\rm nm}$. In this regime, the Casimir force~\cite{casimir1,booknew,french} can cause tiny elements to stick together and cause devices to malfunction\ \cite{Buksall,zhao}.
Tiny elements are also susceptible to wear\ \cite{zhao,William,co4}, thus the operation lifetime of miniaturized machines is a concern. To remedy these problems, it has been
noticed that the lateral Casimir force~\cite{golestan97,Emig-etal-2001,Mohideen,c2010} can intermesh
noncontact parts of nanomachines\ \cite{emig,amg-prl,amg-pre,AFR3,mg-apl,mng-pre,AFR6,nasiri-noise,sync}. 
In view of designing useful nanoscale mechanical devices, the dependence of Casimir force between bodies on their material and geometrical properties is still a subject of intense investigation\ \cite{Vaidya,Lussange,Lombardo,Mohideen2,noncontact gears1,noncontact gears2,bao}.

\begin{figure}[b]
\includegraphics[width=0.7\columnwidth]{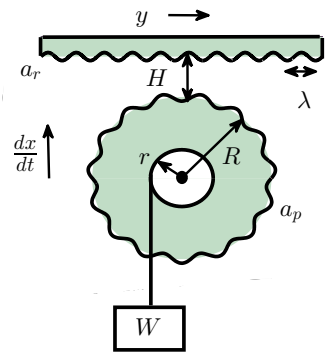}
\caption{(Color online) The Casimir rack and pinion.
The corrugation wavelength $\lambda$, corrugation amplitudes $a_r$ and $a_p$, mean separation $H$, radii $R$ and $r$, external load $W$, and heat bath temperature $T_B$, characterize the system.
} \label{fig1}
\end{figure}

The simplest noncontact nanomachine is composed of one sinusoidally corrugated plate (rack) and one sinusoidally corrugated cylinder (pinion) subject to an external load, see Fig.~\ref{fig1}.
The pinion experiences the lateral Casimir force $F_{\rm lateral}= -F \sin\left[\frac{2 \pi}{\lambda}(x-y)\right] $, where
$\lambda$, $x$, and $y$ denote the corrugation wavelength, pinion displacement, and rack displacement, respectively.
Clearly, the output signal $x(t)$ is a nonlinear function of the input signal $y(t)$. In other words, the Casimir rack and pinion is a {nonlinear} dynamical system. Inspired by the electronic rectifier based on the diode,
a {\it mechanical rectifier} based on the Casimir rack and pinion, has been proposed.
It has been shown that the pinion rotates with a nonzero average velocity and works against an external load,
whether the rack position versus time is a uniform\ \cite{amg-prl}, sinusoidal\ \cite{amg-pre}, or periodic triangular signal\ \cite{AFR3}.

The Casimir rack and pinion is not a perfect mechanical rectifier.
Firstly, it is not guaranteed that an arbitrary input signal can be rectified.
In other words, an arbitrary motion of the rack does not necessarily leads to an {\it upward} motion of the load.
Secondly, the device performance may be degraded due to the thermal noise.
Note that the pinion is mounted on an axle which is surrounded by a fluid (air).
The dynamics of such a small pinion with negligible inertia is deeply influenced by the continuous bombardment of fluid molecules. According to the classic work of Langevin, the total torque on axle can be written as sum of
a frictional torque and a random torque. The fluctuation-dissipation theorem expresses the noise strength in terms of the rotational friction coefficient and the temperature of the system.

Here we study the effect of thermal noise on the strongly damped rack and pinion
device. We assume that the rack position versus time is a periodic multiharmonic signal
\begin{equation}
y(t)=  \sum_{n=1}^{N_h} a_n \sin (n \omega t + \varphi_n),  \label{fo}
\end{equation}
whose time average is zero. Indeed by a proper choice of the number of
harmonics $N_h$, parameters $a_n$ and $\varphi_n$, any signal of period $2 \pi/\omega$ can be well approximated by
Fourier series (\ref{fo}).
We show that in a large domain of parameter space and at room temperature, the rectifier works well.
We observe an interesting phenomenon:
The highly damped rack and pinion does not rectify the celebrated signal $y(t)= a_1 \sin (\omega t +\varphi_1) $, but
many other multiharmonic signals. Indeed the device may not operate for a certain rack motion
whose highest frequency is $ N_h \omega$, but upon adding terms with a {\it higher} harmonic
$ N'_h \omega > N_h \omega $ to that motion, the device lifts {\it up} the load.
The practical message is clear: There is no advantage in reducing the number of harmonics of the input signal.

\section{The Langevin Equation of Motion}\label{sec:eq}

The Casimir rack and pinion shown schematically in Fig.~\ref{fig1},
can be characterized with the corrugation amplitudes $a_p$ and $a_r$, corrugation wavelength $\lambda$,
mean separation $H$, length of the cylinder $L$, radius of the cylinder $R$, radius $r$,
and external load $W$. A heat bath at temperature $T_B$ is in contact with the system.

The lateral Casimir force exerted on the pinion reads $F_{\rm lateral}= -F \sin\left[\frac{2 \pi}{\lambda}(x-y)\right] $, where
$x$ and $y$ are the lateral displacement of the pinion and the rack, respectively.
The amplitude of the lateral Casimir force $F$ depends on the parameters $H$, $\lambda$, $a_p$, $a_r$, $R$, $L$, and $T_B$~\cite{golestan97,Emig-etal-2001,amg-prl,jalaltemperature}.

To describe the dynamics of the heavily damped pinion, we rely on the following Langevin equation
\begin{equation}
-\frac{\zeta}{R}  \frac{d x}{dt }-R F\sin \left[\frac{2\pi}{\lambda}(x-y) \right]- r W + \eta(t)  =0 \label{asli}.
\end{equation}
Here $\zeta$ quantifies the rotational friction. $R F_{\rm lateral}$ and $r W$ are the Casimir torque and the external torque, respectively. The random {torque} $\eta(t)$ is a stationary Gaussian white noise with zero mean and the correlation
$\langle\eta(t_1)\eta(t_2)\rangle=2D\delta(t_1-t_2)$. According to the fluctuation-dissipation theorem $D= \zeta k_B T_B $, where $k_B$ is the Boltzmann's constant.

We choose $ \lambda/(2 \pi)$, ${\zeta \lambda}/{(2\pi R^2 F)}$, and $ V_S={R^2  F}/{\zeta} $
as the units of length, time, and velocity, respectively. We let starred quantities to denote dimensionless variables, e.g.
$x^*=2 \pi x/\lambda$ and $t^*= 2\pi R^2 F t/(\zeta \lambda) $.
The Langevin equation (\ref{asli}) thus can be written as
\begin{equation}
\frac{dz^*}{dt^*}= g(z^*,t^* )+ \eta^*(t^*) \label{moaser}
\end{equation}
where
$z^* \! =\! x^*-y^* $ and
\begin{equation}
 g(z^*,t^* )= - \frac{dy^*}{dt^*}  -\sin z^*  - W_S.  \label{myg}
\end{equation}
$\eta^*(t^*)$ is a Gaussian noise with zero mean and the correlation
$\langle\eta^*(t_1^*)\eta^*(t_2^*)\rangle=2D_S \delta(t_1^*-t_2^*)$. Here the dimensionless parameter
\begin{equation}
D_S=  \frac{2 \pi D }{\zeta \lambda F }= \frac{2 \pi k_B T_B}{\lambda F}, \label{myds}
\end{equation}
compares the noise strength and the Casimir grip.
The dimensionless parameter
\begin{equation}
W_S=\frac{ W r}{ F R }
\end{equation}
compares the external torque and the Casimir torque.

We have assumed that the rack undergoes periodic motion
$y(t)= \displaystyle \sum_{n=1}^{N_h} a_n \sin (n \omega t + \varphi_n)$. It immediately follows that
$ g(z^*,t^* )$ is a space and time periodic function
\begin{equation}
g(z^* + 2 \pi,t^* )= g(z^*,t^* )= g(z^*,t^* +    \frac{2 \pi}{\omega^*}  ), \label{gg}
\end{equation}
where $ \omega^*= \zeta \lambda \omega/(2\pi R^2 F) $. Fourier expansion of  $g(z^*,t^* )$ reads
\begin{equation}
g(z^*,t^* ) =  \sum_{p=-1}^{p=+1}       \sum_{q=-N_h}^{q=+N_h}  g_{p,q} e^{ i p z^* + i q \omega^* t^* }.  \label{expan1}
\end{equation}

\section{ The Fokker-Planck Equation}

We access the probability distribution $\mathcal{W}(z^*,  t^* )$ via the Fokker-Planck equation
\begin{equation}
\frac{\partial   \mathcal{W}(\! z^*, \! t^* ) }{\partial t^*}  \!=\!
D_S  \frac{ \partial ^2   \mathcal{W}( \! z^*, \! t^* )  }{\partial {z^*}^2 }
- \frac{ \partial  \big(g(z^*, \! t^* )  \mathcal{W}( \! z^*, \! t^* )   \big)    }{\partial z^*} .   \label{fok}
\end{equation}
As stated before, $g(z^*, t^* )$ is invariant under the operations
$ z^* \! \rightarrow \! z^* \!+\! 2 \pi   $ and $  t^* \! \rightarrow \!  t^* \!+\! {2 \pi}/{\omega^*} $.
Thus we seek for a steady state probability distribution $\mathcal{W}(z^*,  t^* )$ with the same symmetries.
Following Denisov, H\"{a}nggi and Mateos~\cite{Denisov}, we use the truncated Fourier expansion
\begin{equation}
\mathcal{W}(z^*,t^* ) =  \sum_{n=-N}^{n=+N}       \sum_{m=-M}^{m=+M}  \mathcal{W}_{n,m} e^{ i n z^* + i m \omega^* t^* } \label{expan2}
\end{equation}
to approximate  $\mathcal{W}(z^*,  t^* )$.
The precision of this approximation can be controlled by the parameters $N$ and $M$.

The probability distribution $\mathcal{W}(z^*,  t^* )$ is normalized, i.e.
$ \int_{0}^{ 2 \pi} \mathcal{W}(z^*,t^* ) dz^* =1  $. This implies that $ \mathcal{W}_{0,0}=1/(2 \pi) $ and
$ \mathcal{W}_{0,m}=0 $ if $ m \neq 0$.
The rest of coefficients can be found by inserting Eqs.~(\ref{expan1}) and  (\ref{expan2}) into (\ref{fok}). This yields the following set of {\it linear} algebraic equations~\cite{Denisov}
\begin{equation}
(i m \omega^*   \!\!+\! D_S n^2  )   \mathcal{W}_{n,m}  \!+\! i n  \!
 \sum_{p=-1}^{p=+1}   \sum_{q=-N_h}^{q=+N_h} g_{p,q}  \mathcal{W}_{n-p,m-q} \!=\! 0,
\end{equation}
for $\mathfrak{N}=(2N+1)(2M+1)$ coefficients $ \mathcal{W}_{n,m} $.

It is advantageous to introduce the one-to-one index transformation $(n,m) \rightarrow  \mathfrak{n}= 1+ (n+N)(2M+1)+ m+M$ and then utilizing standard numerical techniques to find single index variables $ \mathcal{W}_\mathfrak{n}$.
We frequently encounter band diagonal set of linear equations, where nonzero elements of the
$\mathfrak{N} \times \mathfrak{N}$ matrix are along a few diagonal lines adjacent to the main diagonal.
To access the solution more rapidly, we use algorithms optimized for band diagonal equations\ \cite{NR}.

\begin{figure*}[t]\begin{center}
    \begin{minipage}[b]{0.32\textwidth}\begin{center}
        \includegraphics[width=\textwidth]{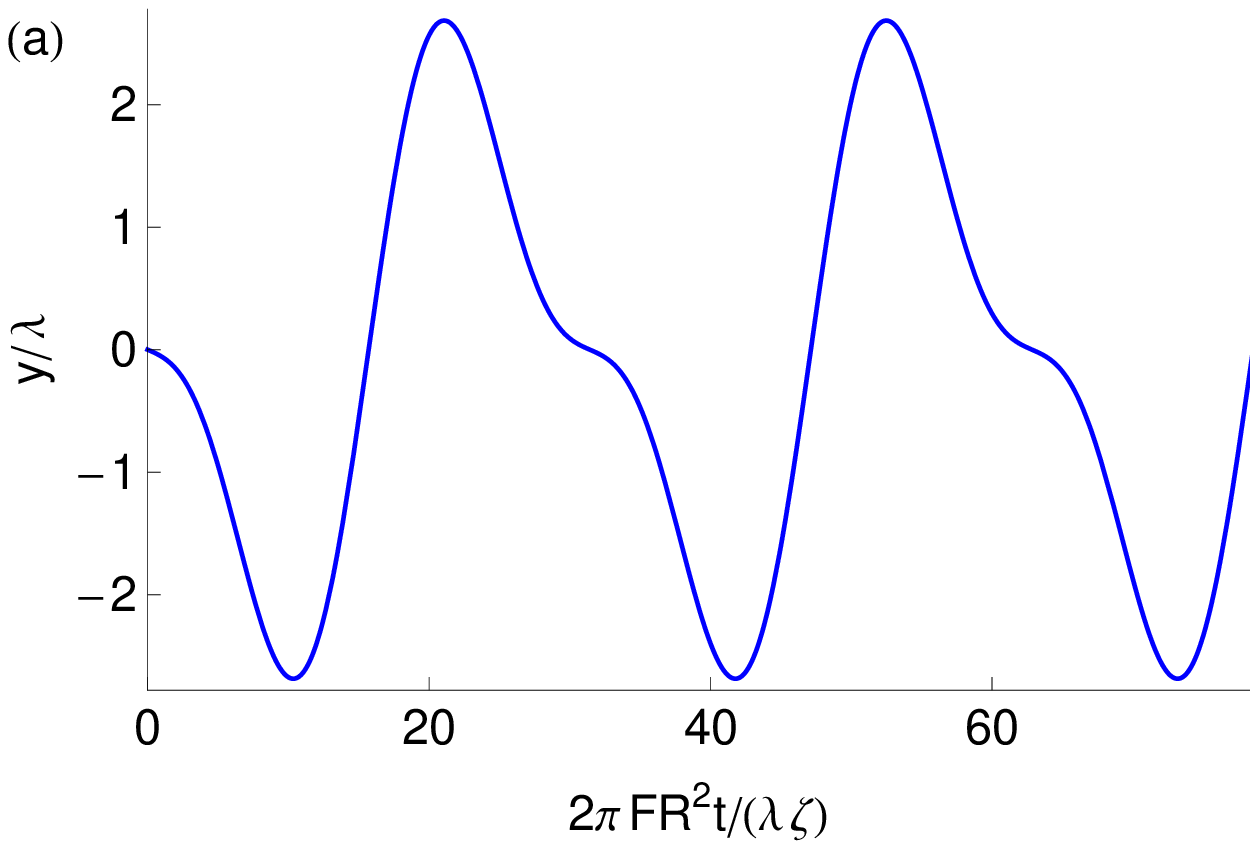}
    \end{center}\end{minipage} \hskip+0cm
    \begin{minipage}[b]{0.32\textwidth}\begin{center}
        \includegraphics[width=\textwidth]{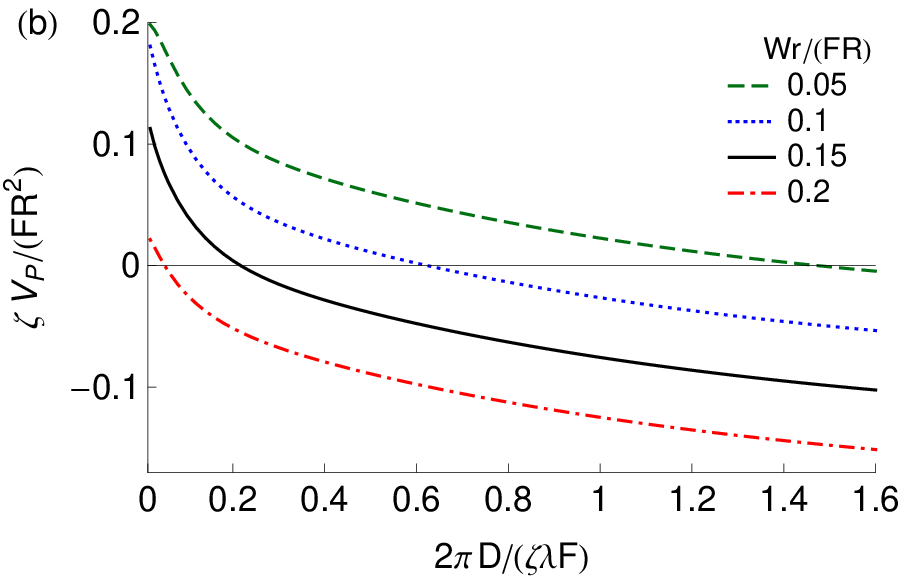}
    \end{center}\end{minipage} \hskip0cm
    \begin{minipage}[b]{0.32\textwidth}\begin{center}
        \includegraphics[width=\textwidth]{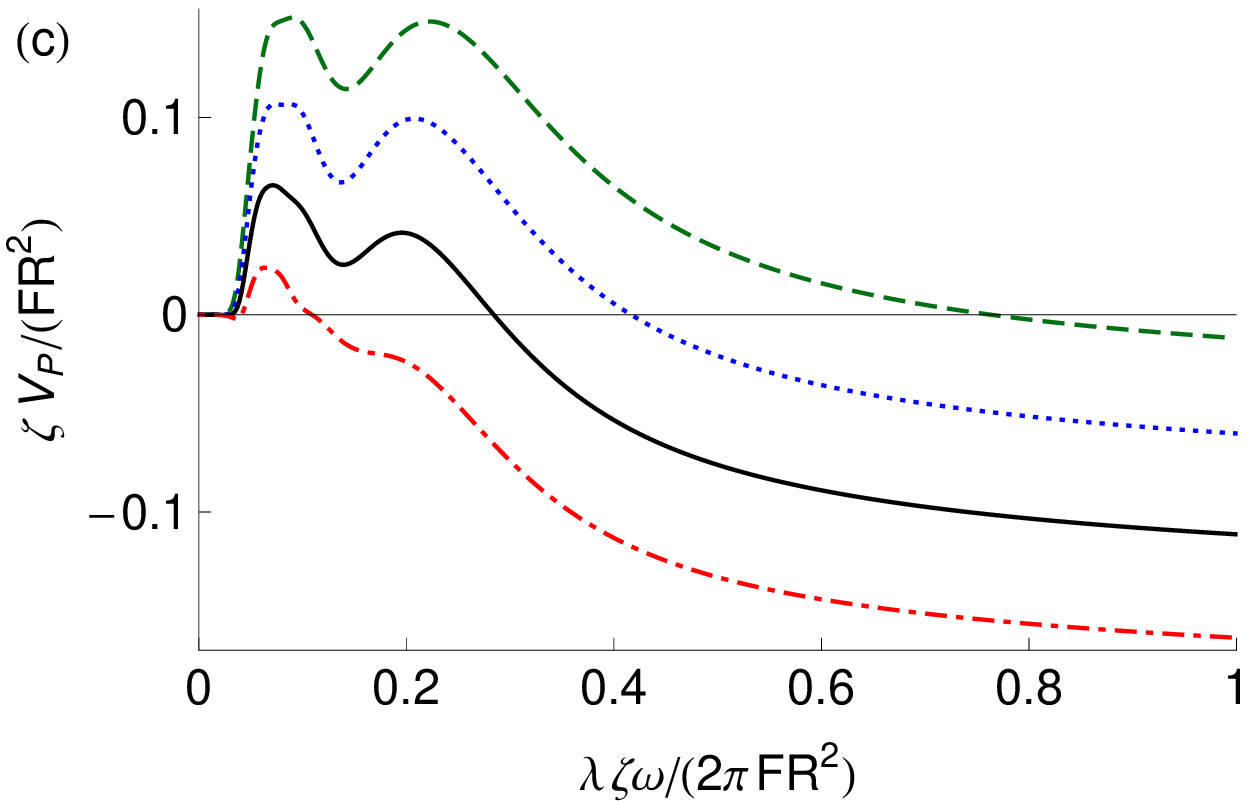}
    \end{center}\end{minipage} \hskip0cm
     \begin{minipage}[b]{0.32\textwidth}\begin{center}
        \includegraphics[width=\textwidth]{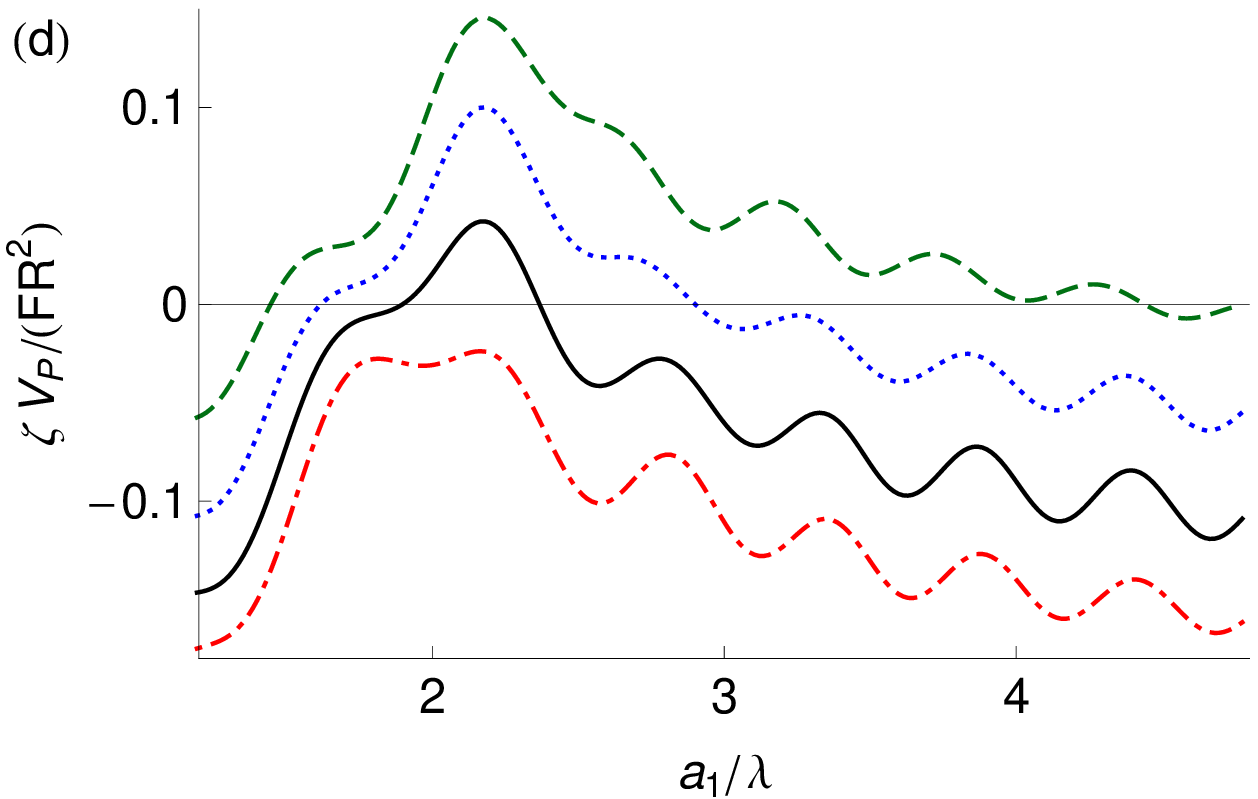}
    \end{center}\end{minipage} \hskip+0cm
    \begin{minipage}[b]{0.32\textwidth}\begin{center}
        \includegraphics[width=\textwidth]{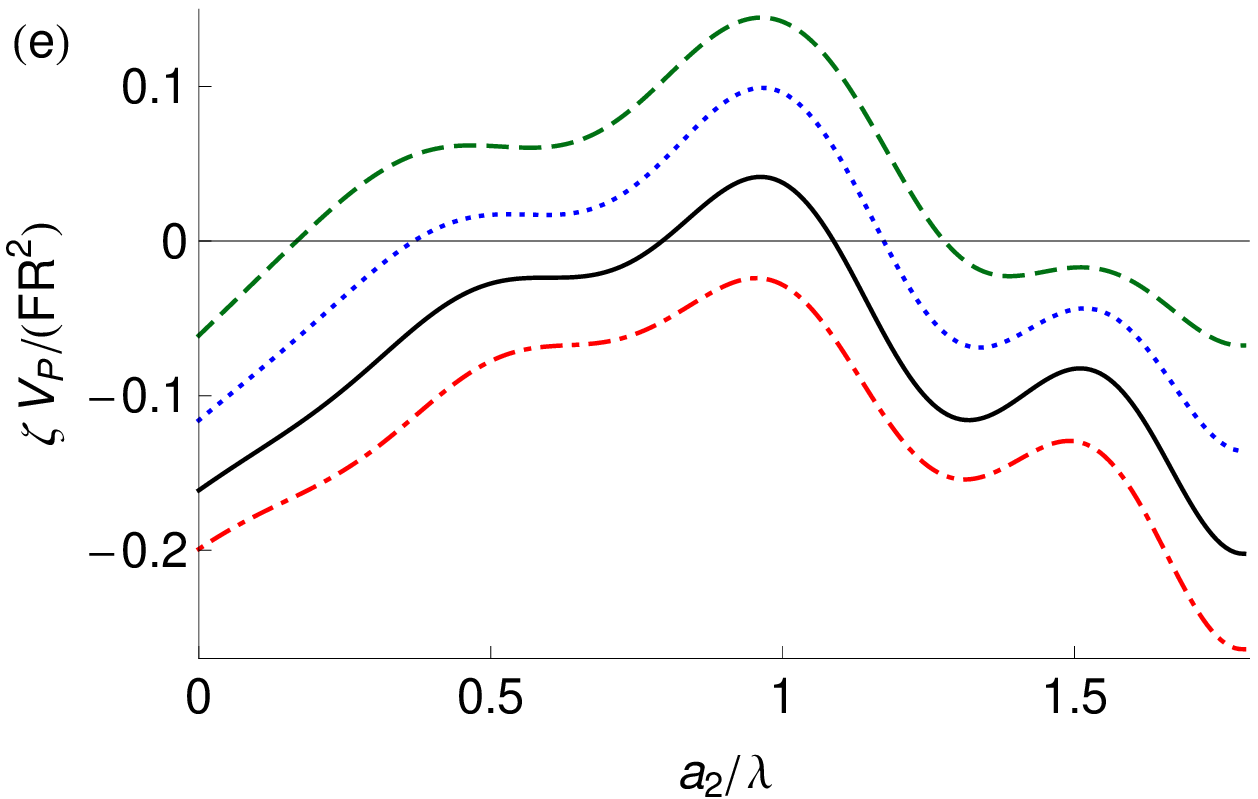}
    \end{center}\end{minipage} \hskip0cm
    \begin{minipage}[b]{0.32\textwidth}\begin{center}
        \includegraphics[width=\textwidth]{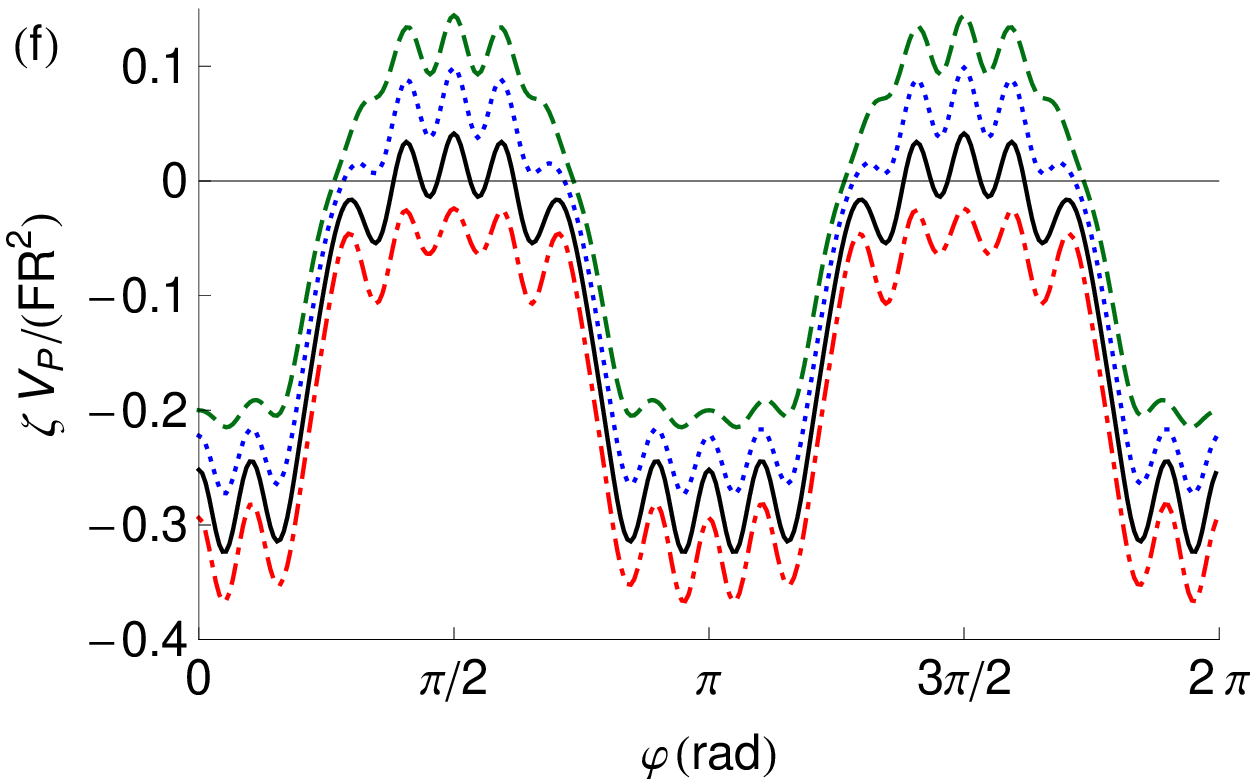}
    \end{center}\end{minipage} \hskip0cm
\caption{(Color online) (a)
The rack position in units of $\lambda$, versus time in units of $\zeta \lambda/(2\pi R^2 F)$, given by Eq.~(\ref{twotwo}).
The average pinion velocity $V_P$ in units of ${R^2  F}/{\zeta}$ versus (b) $2 \pi D /(\zeta \lambda F)$,
(c) $\zeta \lambda \omega/(2\pi R^2 F)$, (d) $a_1/\lambda$, (e) $a_2/\lambda$ and (f) $\varphi$.
Except the parameter being varied, $ 2 \pi a_1/\lambda   \!=\! 13.5$, $ 2 \pi a_2/\lambda \!=\! 6$,
$ \zeta \lambda \omega/(2\pi R^2 F) \!=\!0.2$,  $\varphi \!=\!\pi/2$, and $2 \pi D /(\zeta \lambda F) \!=\!0.1$
are assumed.  } \label{figbi1}
\end{center}
\end{figure*}

The average of the random torque $\eta^*(t^*)$ is zero, thus from Eq. (\ref{moaser}) we find that
$  \langle {dz^*}/{dt^*}   \rangle  =   \langle g(z^*,t^* )   \rangle   $.
Powered by the probability distribution $\mathcal{W}(z^*,  t^* )$, we find
\begin{eqnarray}
  \langle \frac{dz^*}{dt^*}   \rangle \!&=& \!\langle g(z^*,t^* )   \rangle  ,  \nonumber \\
   & \!\! =\!\! & \lim_{k \rightarrow \infty}  \frac{\omega^*}{2 \pi k}
\int_0^{ \frac{2 \pi k}{\omega^*}}    \! \! \int_0^{2 \pi} g(z^*,t^* )   \mathcal{W}(z^*,t^* )  dz^*  dt^*  , \nonumber \\
 &\!\! =\!\! &    2 \pi   \sum_{p=-1}^{p=+1}   \sum_{q=-N_h}^{q=+N_h} g_{p,q}  \mathcal{W}_{-p,-q}.
\end{eqnarray}
Recalling that the average rack velocity is zero, we find the average pinion velocity
\begin{equation}
V_P= \langle {dx}/{dt}   \rangle=   2 \pi V_S   \sum_{p=-1}^{p=+1}   \sum_{q=-N_h}^{q=+N_h} g_{p,q}  \mathcal{W}_{-p,-q}  ,
\end{equation}
in terms of the easily accessible Fourier coefficients $\mathcal{W}_{n,m}$.
This key result allows us to fully investigate the rectifier performance.

\begin{figure*}[t]\begin{center}
    \begin{minipage}[b]{0.32\textwidth}\begin{center}
        \includegraphics[width=\textwidth]{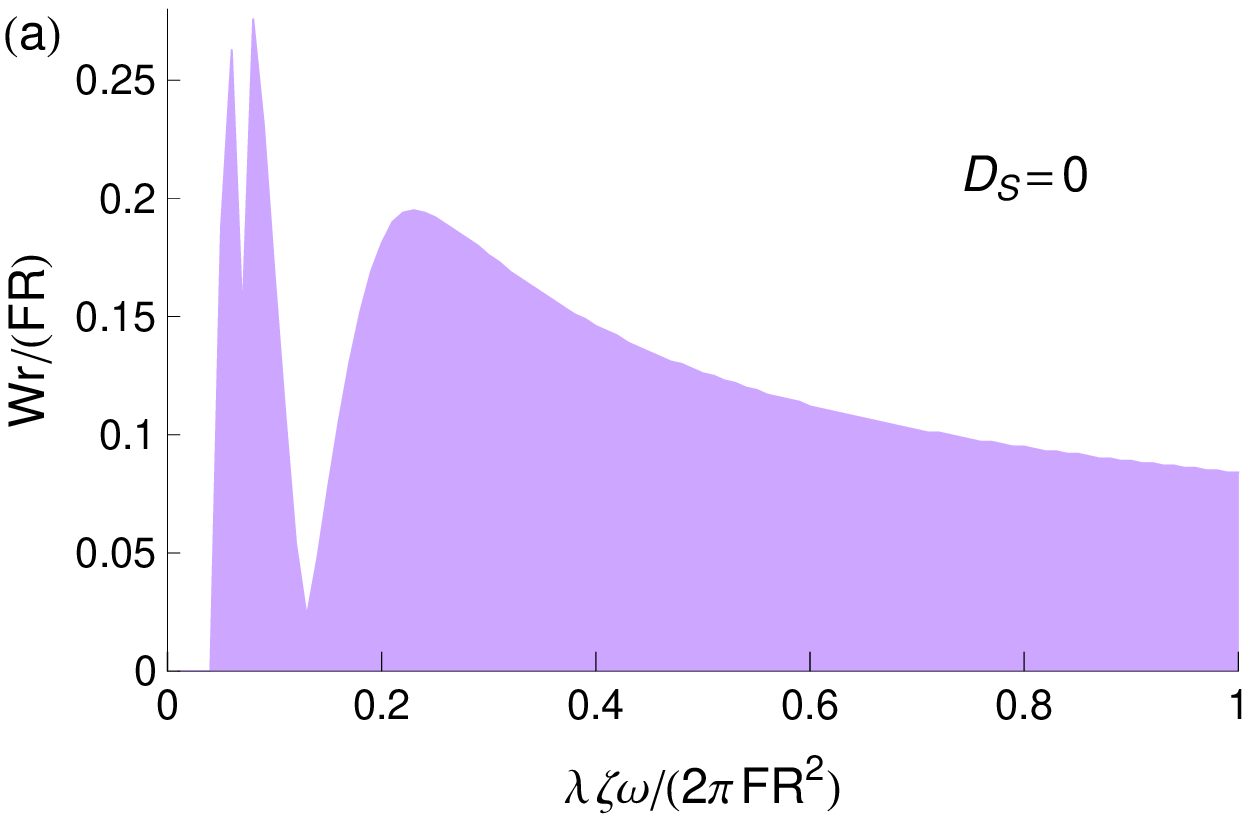}
    \end{center}\end{minipage} \hskip+0cm
        \begin{minipage}[b]{0.32\textwidth}\begin{center}
        \includegraphics[width=\textwidth]{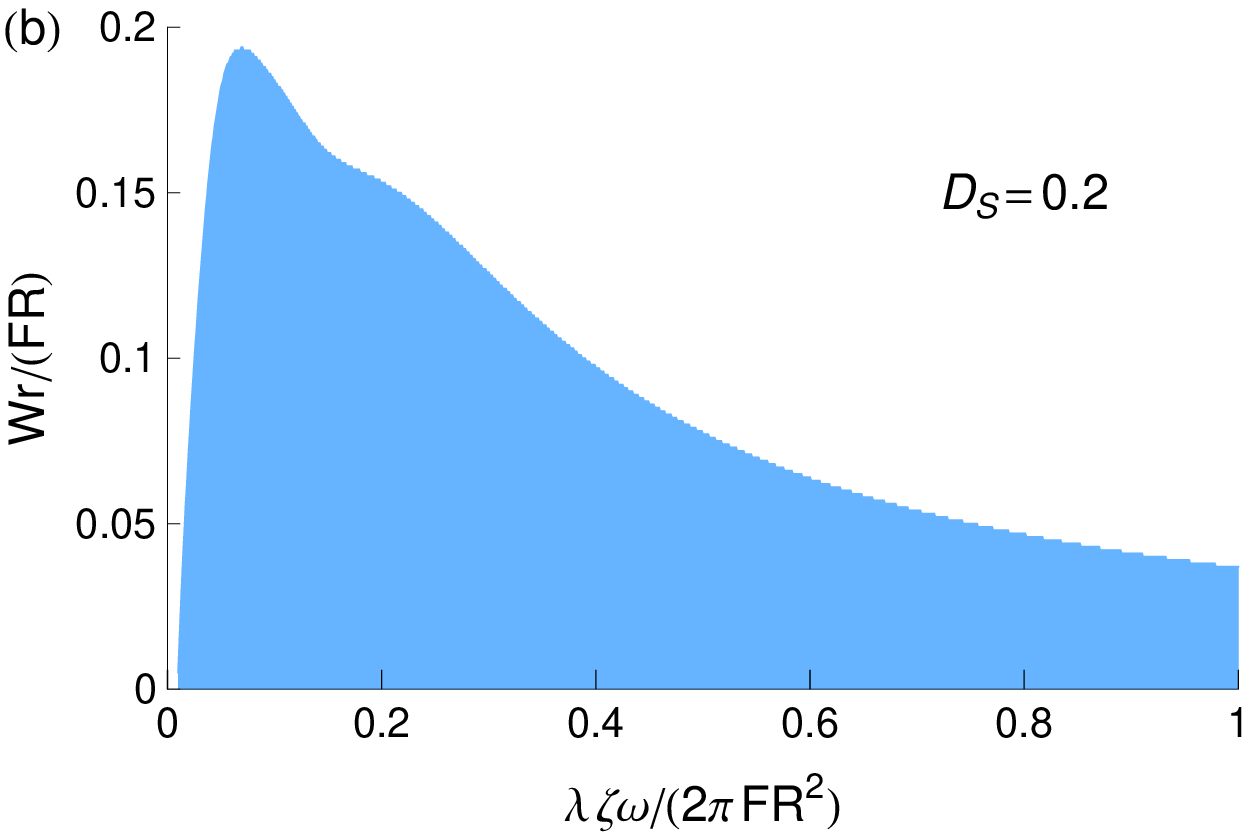}
    \end{center}\end{minipage} \hskip0cm
            \begin{minipage}[b]{0.32\textwidth}\begin{center}
        \includegraphics[width=\textwidth]{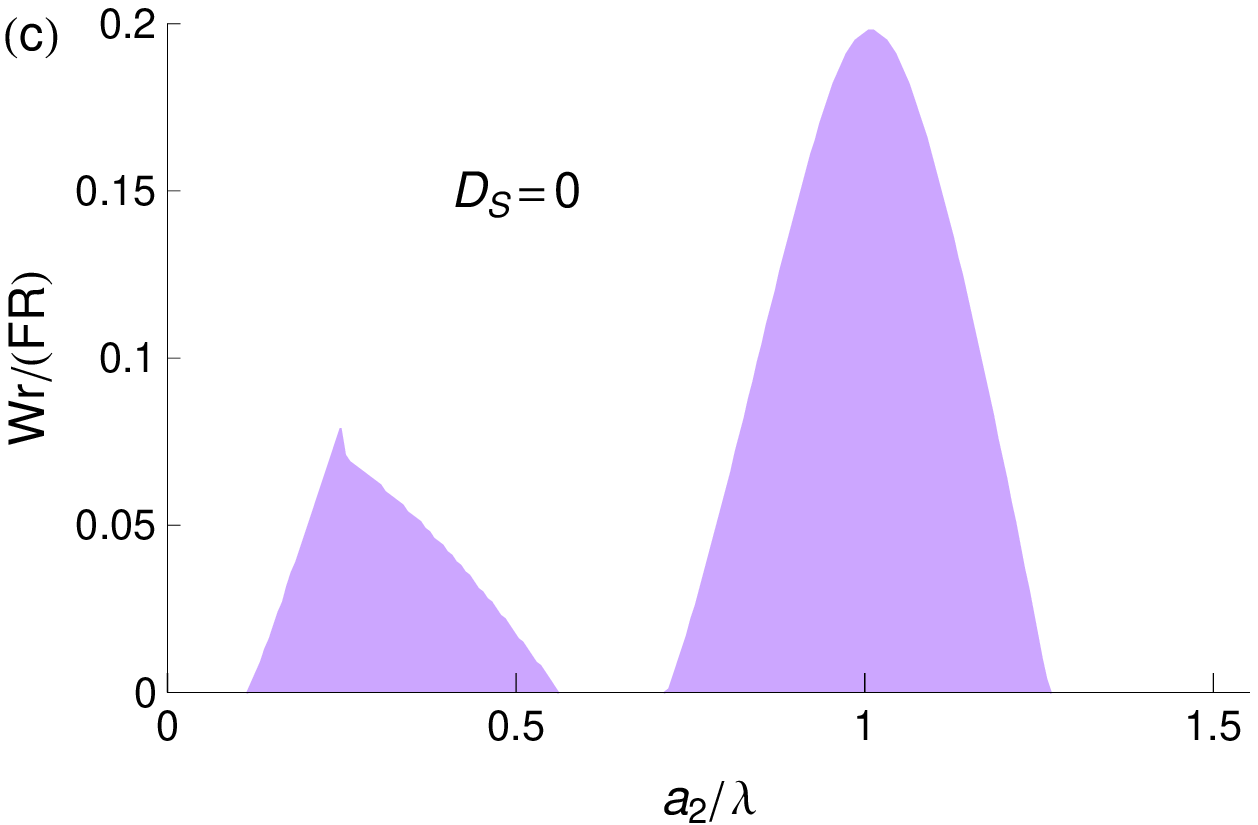}
    \end{center}\end{minipage} \hskip0cm
    \begin{minipage}[b]{0.32\textwidth}\begin{center}
        \includegraphics[width=\textwidth]{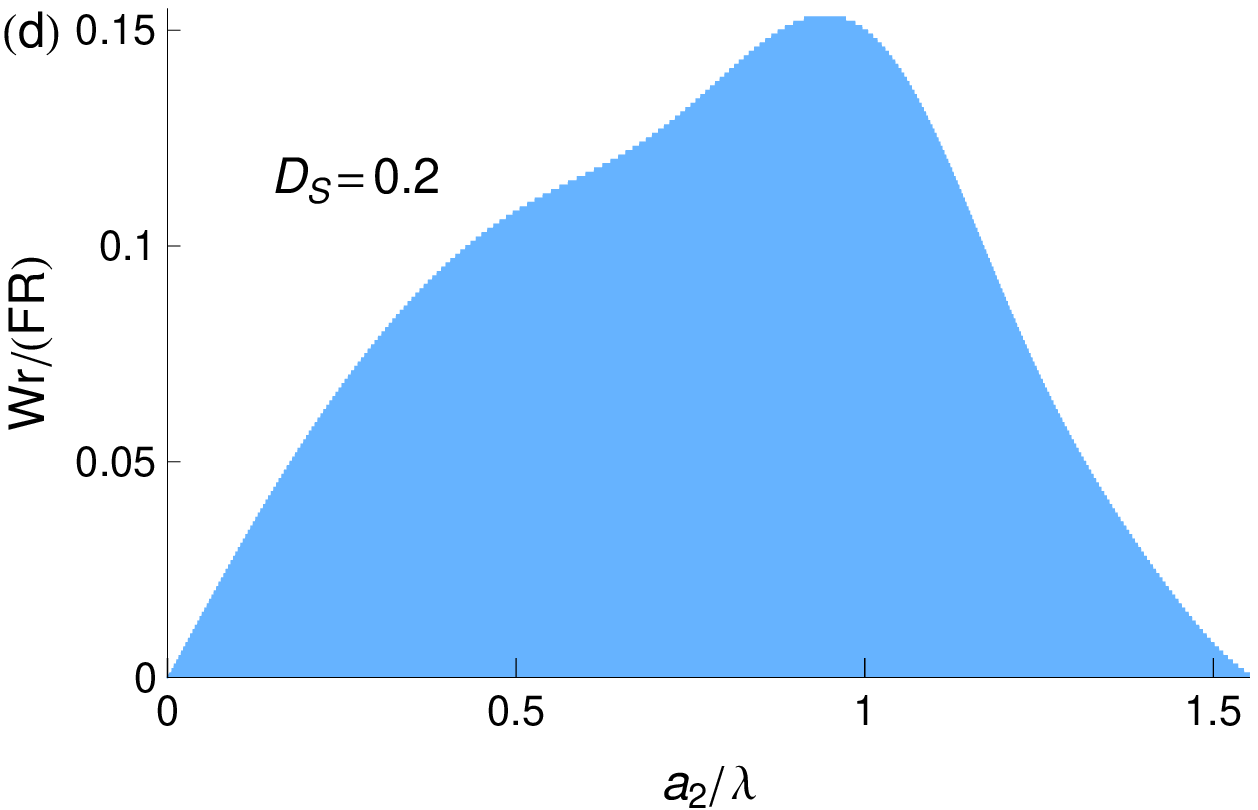}
    \end{center}\end{minipage} \hskip0cm
    \begin{minipage}[b]{0.32\textwidth}\begin{center}
        \includegraphics[width=\textwidth]{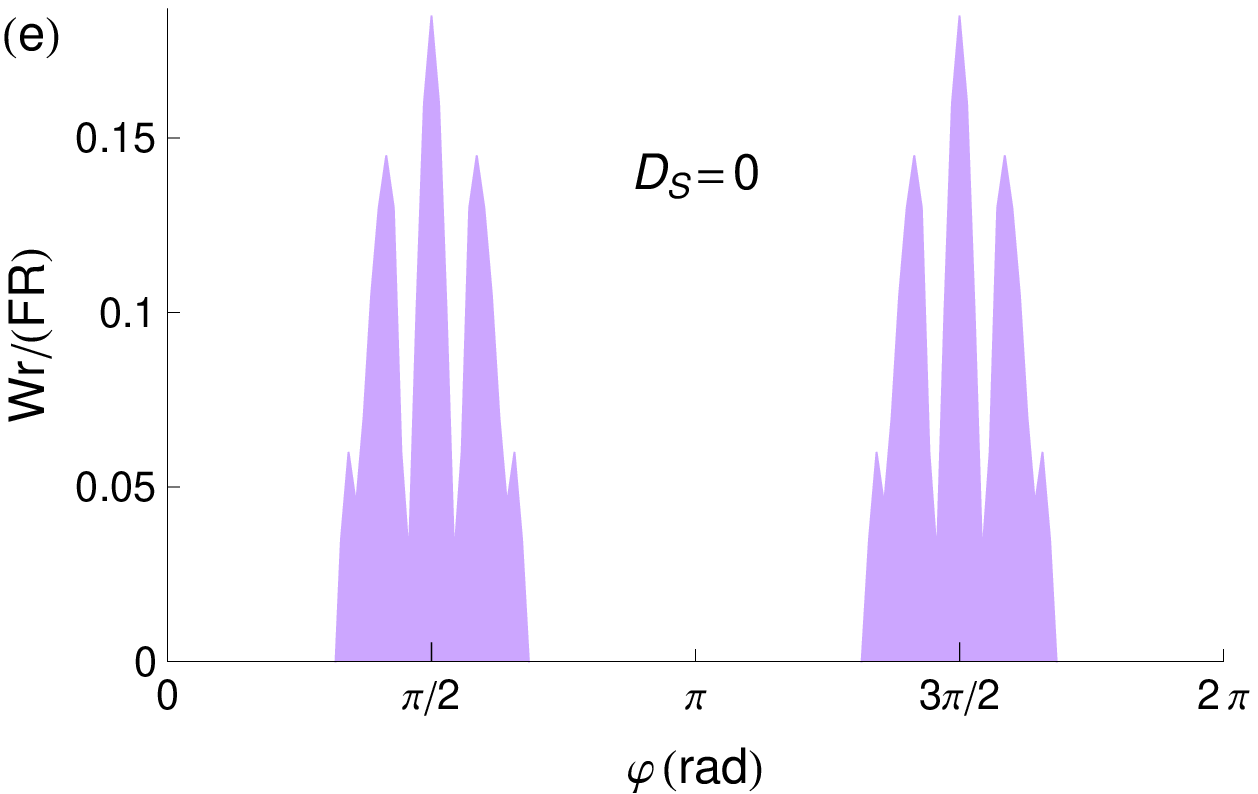}
    \end{center}\end{minipage} \hskip0cm
    \begin{minipage}[b]{0.32\textwidth}\begin{center}
        \includegraphics[width=\textwidth]{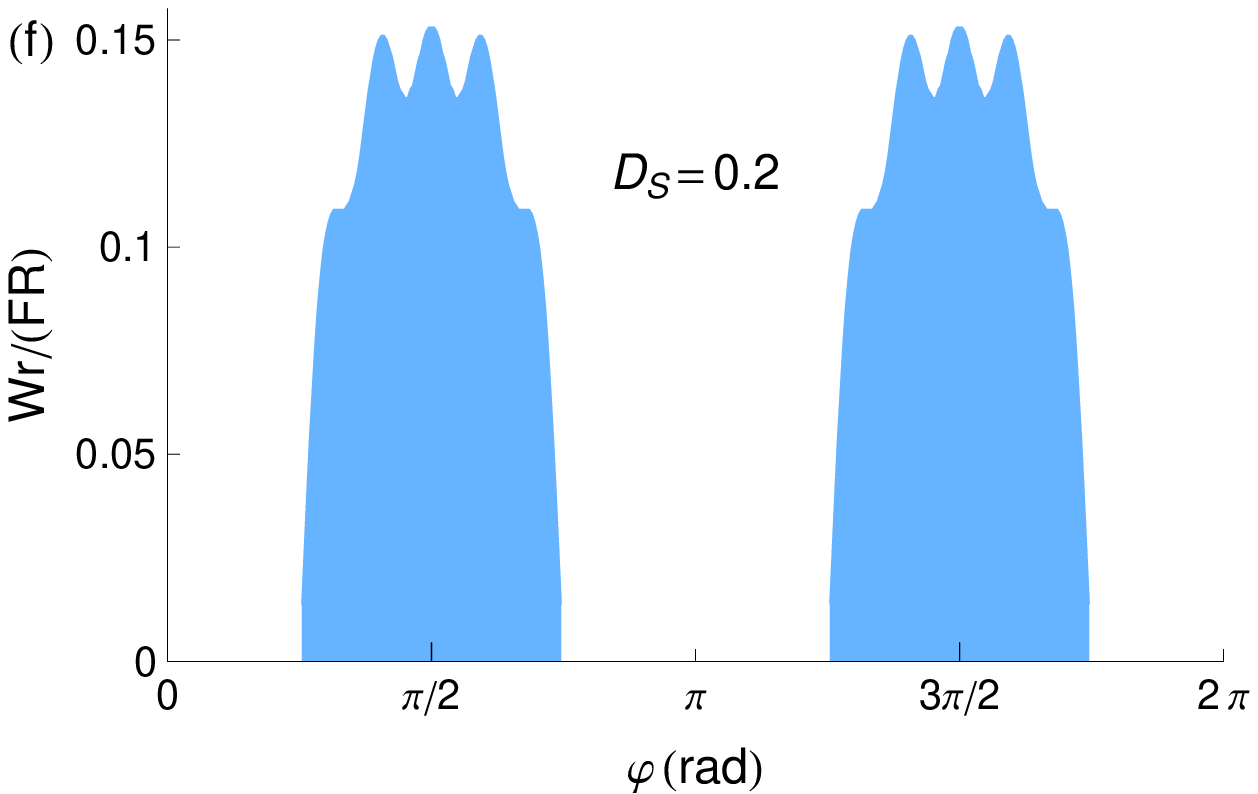}
    \end{center}\end{minipage} \hskip0cm
\caption{(Color online) Domain of positive pinion velocities in the (a) and (b) $ (W_S, \omega^*) $,
(c) and (d) $ (W_S, a_2/\lambda ) $,  (e) and (f) $ (W_S, \phi) $ parameter space,
for noise strengths $ 2 \pi D /(\zeta \lambda F) \!=\!0$ and $ 2 \pi D /(\zeta \lambda F) \!=\!0.2$.
Except the parameter being varied, $ 2 \pi a_1/\lambda   \!=\! 13.5$, $ 2 \pi a_2/\lambda \!=\! 6$,
$ \zeta \lambda \omega/(2\pi R^2 F) \!=\!0.2$, and $\varphi \!=\!\pi/2$ are assumed.} \label{figbi2}
\end{center}
\end{figure*}

\section{Typical values of parameters}\label{typical}
Inspired by the experimental setup of Chen {\it et al.}\ \cite{Mohideen}, we assume $\lambda=500 \;{\rm nm}$, $R=1 \;\mu{\rm m}$, and $a_r=a_p=50 \;{\rm nm}$. At $ T_B=0$, we estimate $F=0.3 \;{\rm pN}$ for $H=200 \;{\rm nm}$
and $F=11.7 \;{\rm pN}$ for $H=100 \;{\rm nm}$~\cite{amg-prl}.
For these values of $\lambda$, $R$, and $H \ll   \lambda_T= \hbar c /( k_B T_B) $, the amplitude of the lateral Casimir force does not change significantly as temperature raises to $T_B=300 \;{\rm K} $~\cite{jalaltemperature}.
Typically, the load $W \!\sim\! 0.1 \!-\!1 \; F$ and $r \!\sim\! R $, thus $W_S \!\sim\! 0.1\!-\!1 $. The rotational friction coefficient $\zeta$ is about $ 8 \times 10^{-20 } \; {\rm kg}\; {\rm m}^2 /{\rm s} $~\cite{mg-apl,AFR3}.
Thus for $F=0.3 \;{\rm pN}$ and $T_B=300 \;{\rm K} $, we find $ V_S=3.75  \;\mu{\rm m}/{\rm s} $ and $D_S= 0.18 $.

Considering the signal $y(t)$, we assume $a_n  \! \sim\!  1\!-\!2  \; \lambda $ and $ \omega \!\sim\! 10-50 \;{\rm Hz}  $, thus
$a^*  \!\sim\! 1-15 $ and $ \omega^*= \zeta \lambda \omega/(2\pi R^2 F) \sim 0.2 \!-\! 1 $.

\section{Biharmonic driving }\label{twoharmonics}

Now we consider the biharmonic signal
\begin{equation}
y(t)=a_1 \sin (\omega t + \varphi) + a_2 \sin (2 \omega t ).  \label{twotwo}
\end{equation}
The dimensionless amplitudes $a_1^*  \!=\! {2 \pi a_1}/{\lambda} $,
$ a_2^*  \!=\!      {2 \pi a_2}/{\lambda} $, dimensionless frequency
$\omega^*    \! =\!  { \zeta \lambda \omega}/{(2\pi R^2 F)}$, and phase $\varphi$ characterize this signal.
As a concrete example, we consider a biharmonic signal with $a_1^*  \!=\! 13.5$,
$a_2^*  \!=\! 6$, $\omega^* \!=\!0.2$, and $\varphi  \!=\!\pi/2$, see Fig.~\ref{figbi1}(a).
In our following study of the average pinion velocity, we change {one} of the parameters and keep other parameters fixed.

Figure~\ref{figbi1}(b) demonstrates $V_P/V_S $ as a function of noise strength and for various loads.
$V_P/V_S$ monotonically decreases as the noise strength increases.
Remarkably, even for the load $ W_S \!=\! 0.15$,
the average velocity is positive for noise strengths $D_S <0.22$.

Hereafter we assume $D_S \!=\! 0.1$. Figure~\ref{figbi1}(c) shows $V_P/V_S$ as a function of $\omega^*$.
Remarkably, $V_P/V_S$ is zero for frequencies smaller than a threshold $\omega^*_{th}$.
Here, almost independent of the load,  $\omega^*_{th}= 0.03$. $V_P/V_S$ is monotonically decreasing when $\omega^*  \!>\! \omega^{*}_{\times}$. For example,
$ \omega^{*}_{\times}= 0.23$ when $ W_S \!=\! 0.05$, but shifts to $ \omega^{*}_{\times}= 0.2$ when $ W_S \!=\! 0.15$.
For $\omega^*< \omega^{*}_{\times}$ and loads $ W_S \! \leqslant \! 0.15$, $V_P/V_S$ exhibits two maxima and one minima. Thus to maximize the average pinion velocity, the frequency must be deliberately tuned.

Figure~\ref{figbi1}(d) shows that $V_P/V_S$ is a {nonmonotonic} function of $a_1/\lambda$.
There exists a window of $a_1/\lambda$ to ensure that the device lifts up the load. Moreover, $a_1/\lambda$ can be tuned
to maximize $V_P/V_S$. In the case of $ W_S \!=\! 0.15$, we find $V_P/V_S$ reaches its maximum $0.04$ at $ a_1/\lambda \!=\! 2.18 $. $V_P/V_S$ is positive in the window $ 1.89 \! <   \! a_1/\lambda  \! < \!2.35 $.
Figure~\ref{figbi1}(e) depicting $V_P/V_S$ as a function of $a_2/\lambda$, conveys similar messages.

Figure~\ref{figbi1}(f) shows the average pinion velocity as a function of the phase.
We find $V_P(\varphi)=V_P(2 \pi -\varphi)$, {\it i.e.}, $V_P/V_S$ is symmetric with respect to the line $\varphi=\pi$.
Moreover, $V_P/V_S$ may be positive in {\it disconnected} windows of $\phi$.
For $ W_S \!=\! 0.15$, these windows are $(1.20,1.37)$, $(1.47,1.60)$, and $(1.76,1.93)$ radians.

One intuitively expects the domain of parameters ensuring $V_P \!> \!0$ to shrink, as the load increases.
Figures~\ref{figbi1}(b)-(f) all confirm this expectation.

In the following, we present a few sections of $(a_1^*,a_2^*,\omega^*, \varphi,W_S, D_S)$ parameter space where the
noisy rack and pinion derived by a biharmonic signal, properly lifts {upward} the external load.
Except the parameter being varied, $a_1^*  \!=\! 13.5$, $a_2^*  \!=\! 6$, $\omega^* \!=\!0.2$, and $\varphi \!=\!\pi/2$
are assumed. We compare the device performance at no noise $D_S=0$, and at room temperature noise $D_S=0.2$.

Figures~\ref{figbi2}(a) and (b) demonstrate domains of positive pinion velocity in the $(W_S,\omega^* )$ space for
$D_S=0$ and $D_S=0.2$, respectively.
At first look, the noise degrades the performance:
For example, at $\omega^*=1$, the maximum load decreases from $W_S=0.084$ to $W_S=0.037$ as the temperature 
increases. But quite remarkably, the noise may facilitate the device operation: For $D_S=0$, a gap $(0, \omega^*_{th}=0.05 )$ and a gap around $\omega^*=0.13$ indicate that the device does not lift up the load.
On increasing the temperature, the first gap shrinks to $(0, \omega^*_{th}=0.01 )$ and the second gap completely disappears.

Figures~\ref{figbi2}(c) and (d) demonstrate domains of positive pinion velocity in the $(W_S, a_2/\lambda)$ space for
$D_S=0$ and $D_S=0.2$, respectively. Focusing on the intervals $ 0 \!<\! a_2/\lambda  \!< \! 0.11 $,
$ 0.56 \!<\! a_2/\lambda \!<\! 0.71 $ and $ 1.27 \!<\! a_2/\lambda \!<\! 1.55 $, again we find that the noise may facilitate the device operation. This conclusion is also supported by Figs.~\ref{figbi2}(e) and (f).

\begin{figure*}[t]\begin{center}
    \begin{minipage}[b]{0.32\textwidth}\begin{center}
        \includegraphics[width=\textwidth]{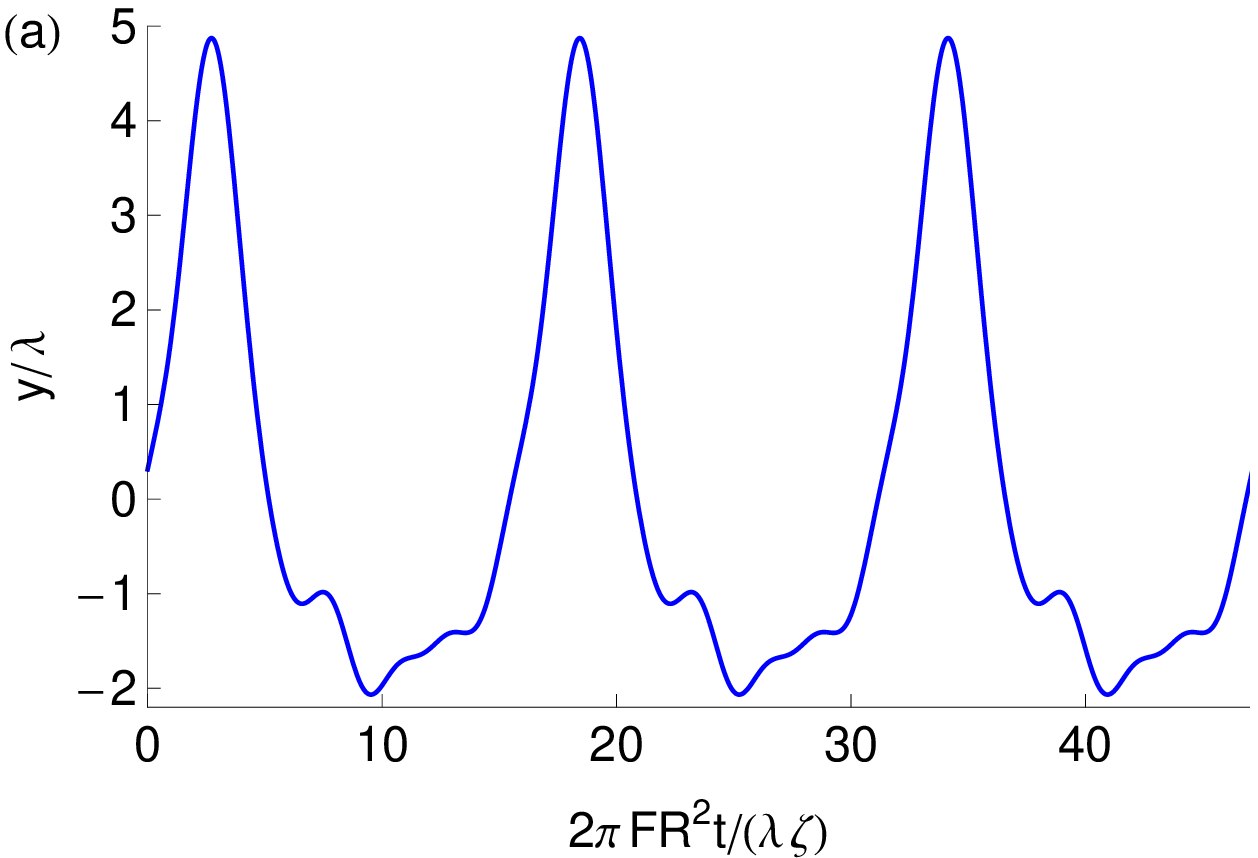}
    \end{center}\end{minipage} \hskip+0cm
    \begin{minipage}[b]{0.32\textwidth}\begin{center}
        \includegraphics[width=\textwidth]{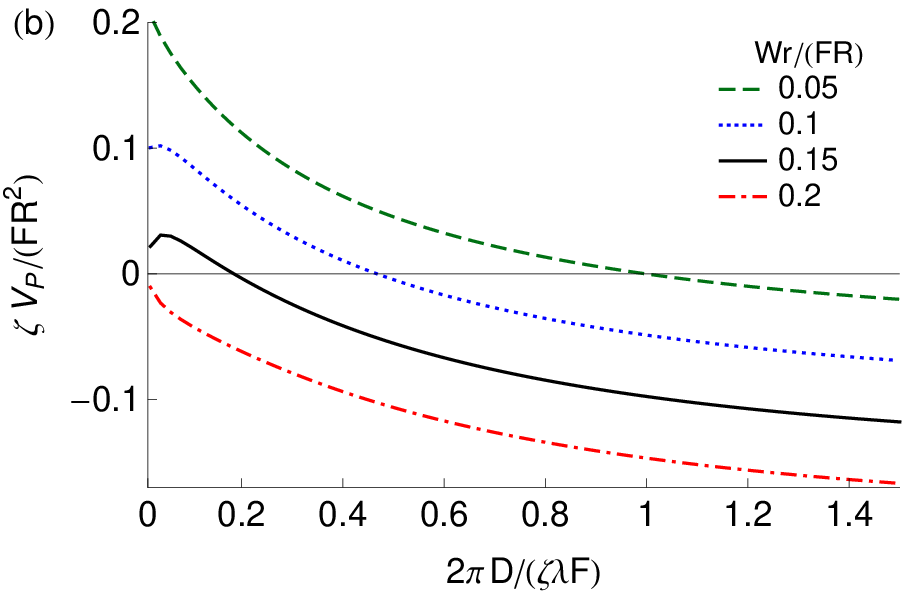}
    \end{center}\end{minipage} \hskip0cm
    \begin{minipage}[b]{0.32\textwidth}\begin{center}
        \includegraphics[width=\textwidth]{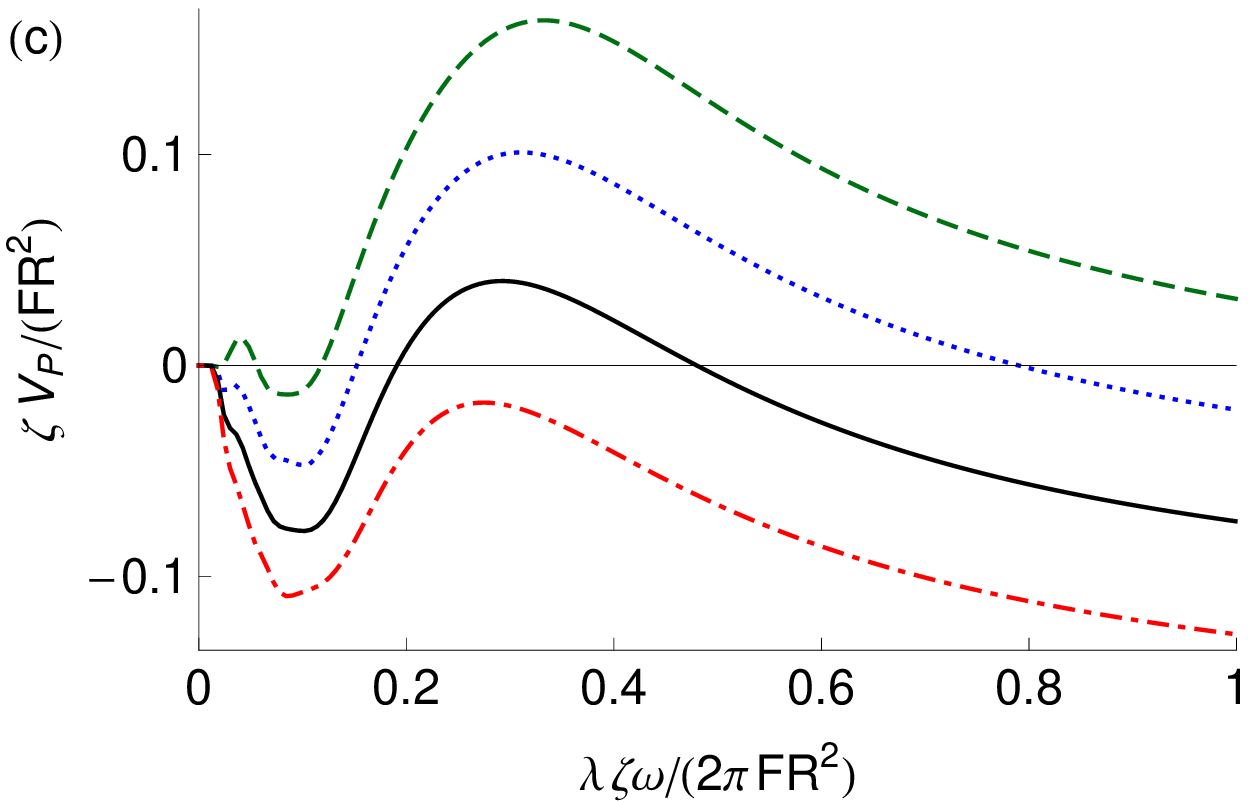}
    \end{center}\end{minipage} \hskip0cm
     \begin{minipage}[b]{0.32\textwidth}\begin{center}
        \includegraphics[width=\textwidth]{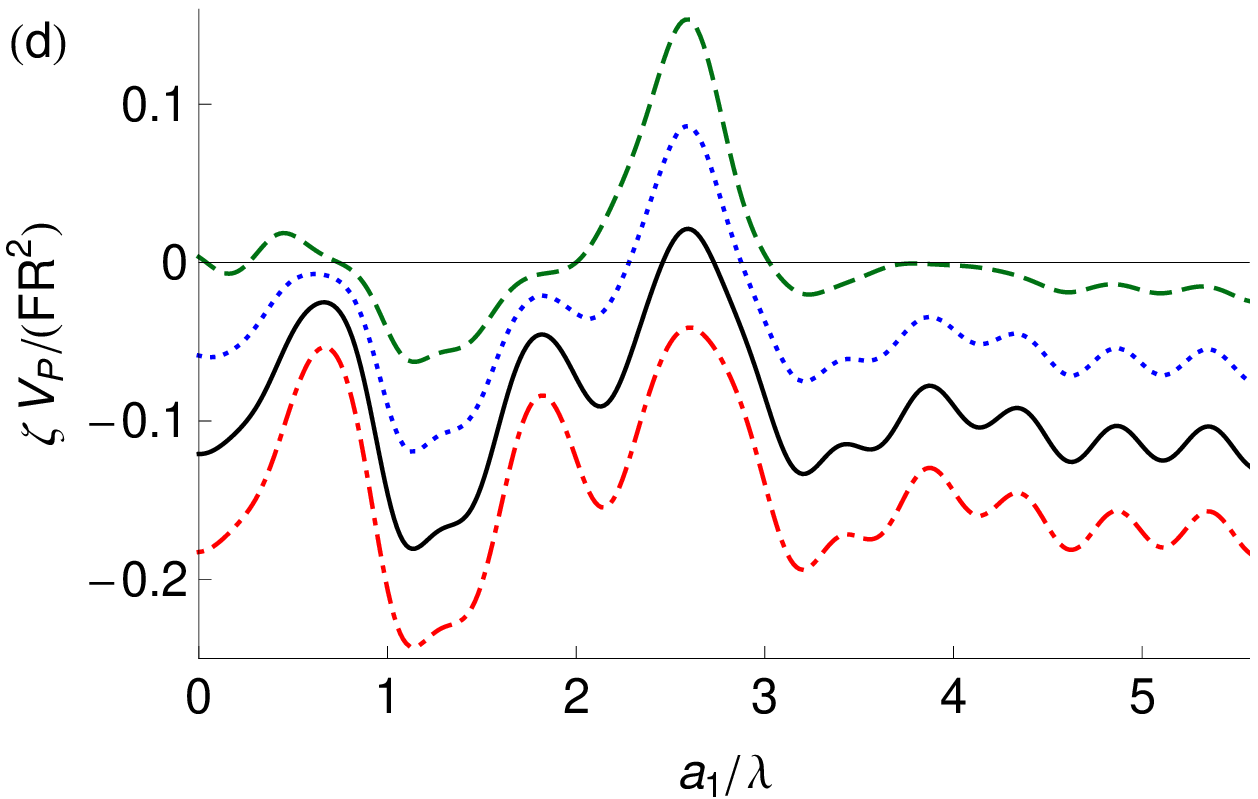}
    \end{center}\end{minipage} \hskip+0cm
    \begin{minipage}[b]{0.32\textwidth}\begin{center}
        \includegraphics[width=\textwidth]{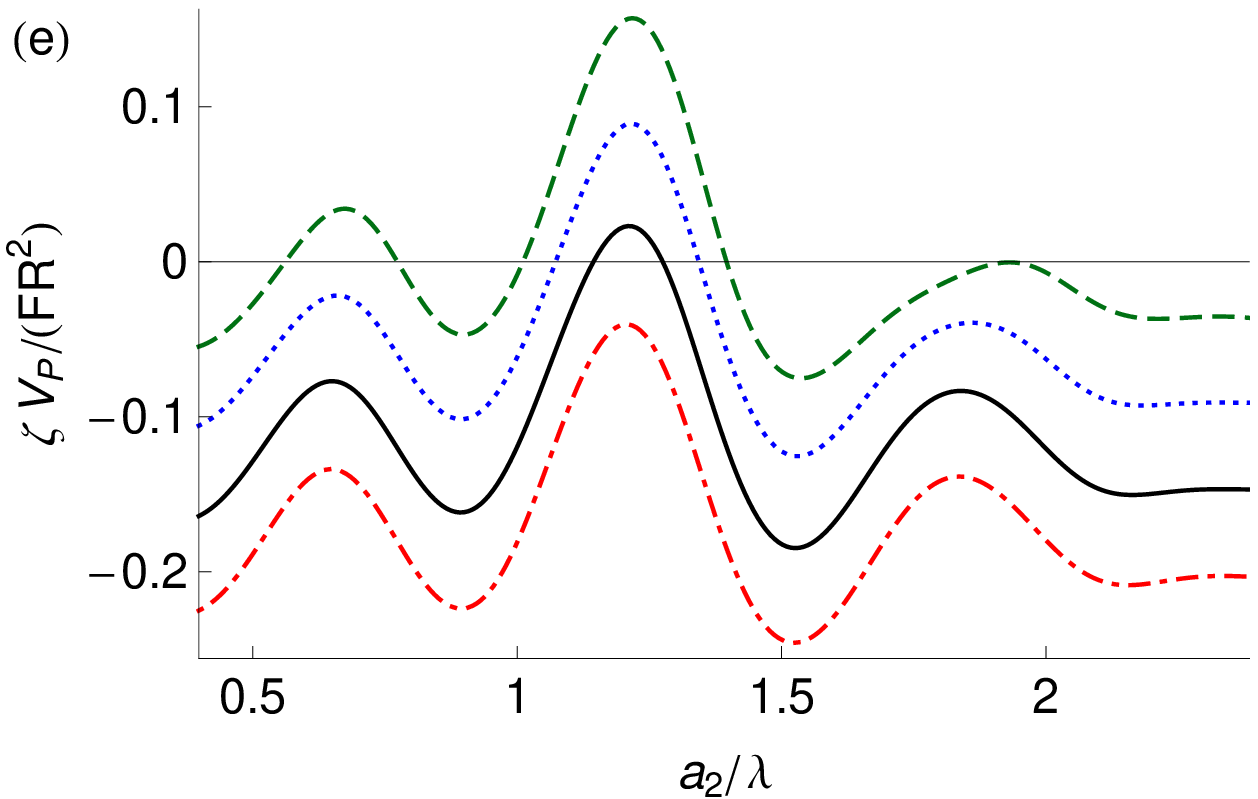}
    \end{center}\end{minipage} \hskip0cm
    \begin{minipage}[b]{0.32\textwidth}\begin{center}
        \includegraphics[width=\textwidth]{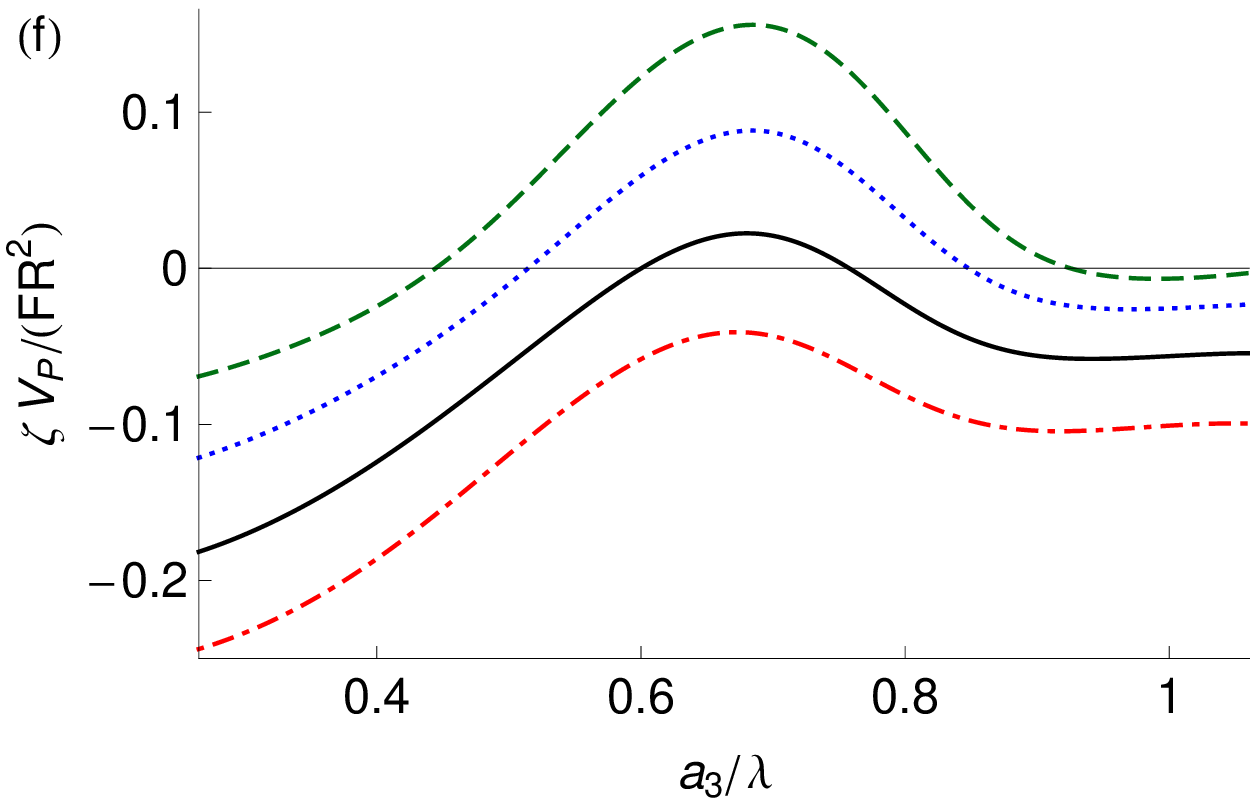}
    \end{center}\end{minipage} \hskip0cm
    \begin{minipage}[b]{0.32\textwidth}\begin{center}
        \includegraphics[width=\textwidth]{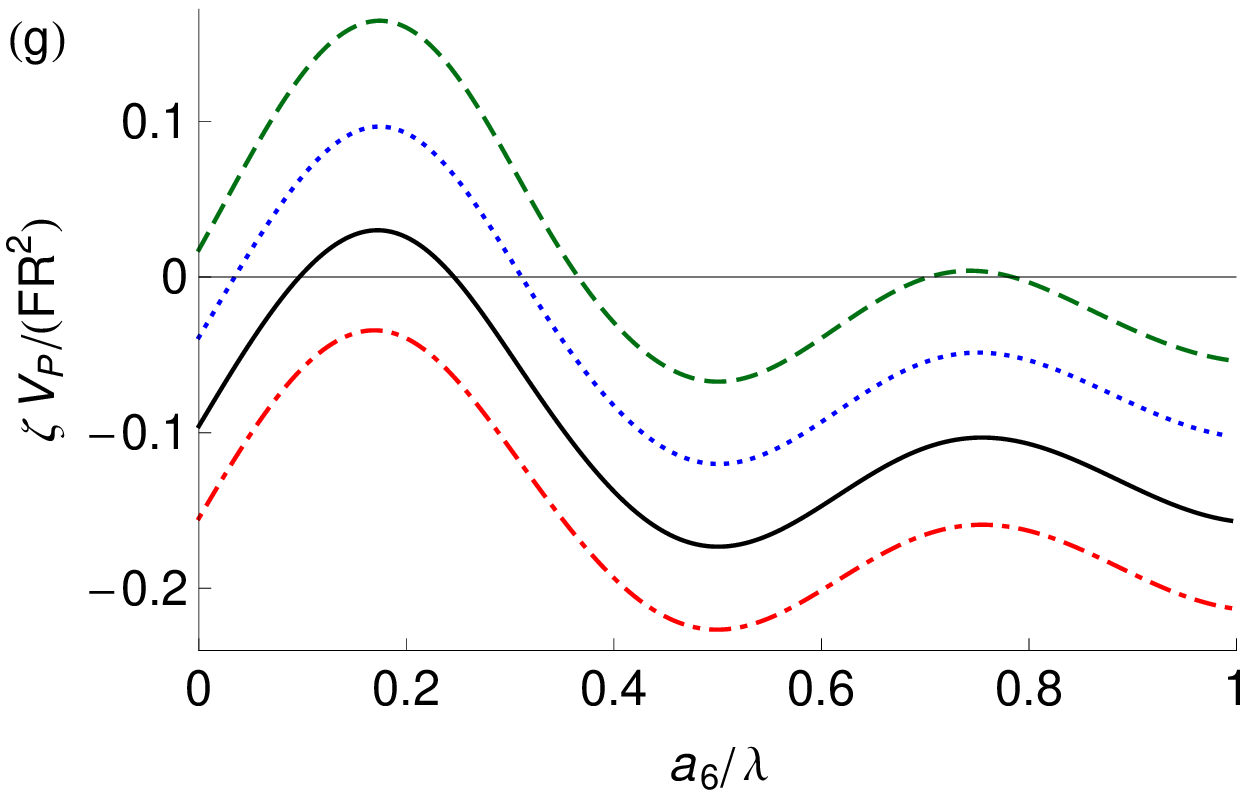}
    \end{center}\end{minipage} \hskip0cm
    \begin{minipage}[b]{0.32\textwidth}\begin{center}
        \includegraphics[width=\textwidth]{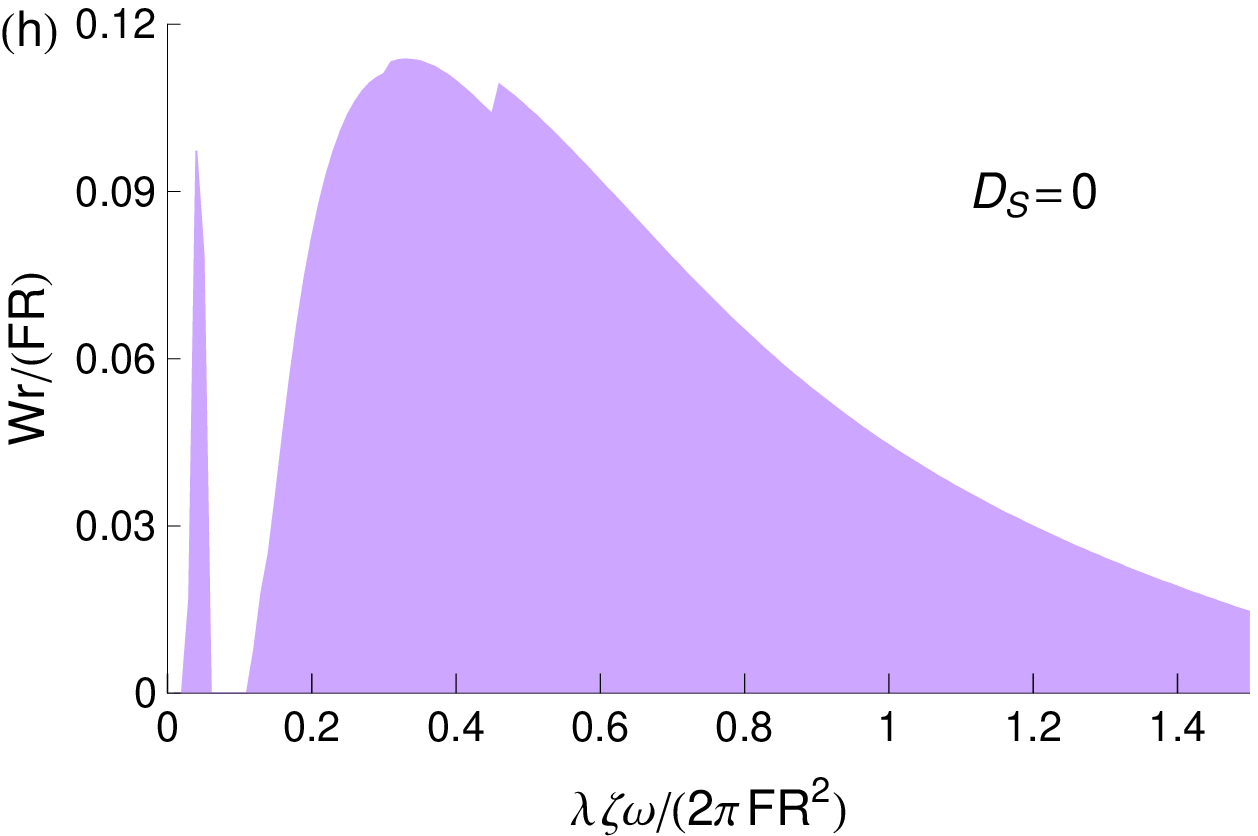}
    \end{center}\end{minipage} \hskip0cm
    \begin{minipage}[b]{0.32\textwidth}\begin{center}
        \includegraphics[width=\textwidth]{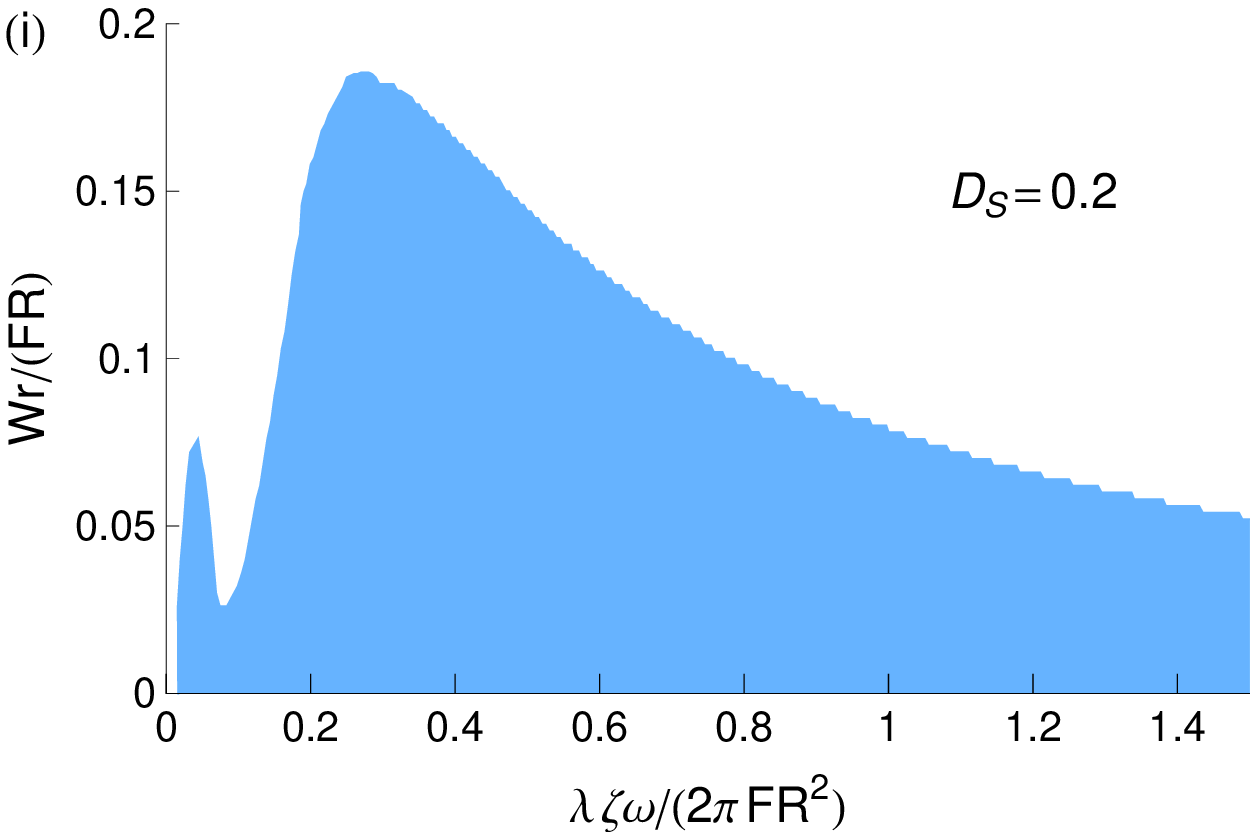}
    \end{center}\end{minipage} \hskip0cm
\caption{(Color online) (a)
The rack position in units of $\lambda$, versus time in units of $\zeta \lambda/(2\pi R^2 F)$, given by Eq.~(\ref{tritri}).
$V_P$ in units of ${R^2  F}/{\zeta}$ versus (b) $2 \pi D /(\zeta \lambda F)$,
(c) $\zeta \lambda \omega/(2\pi R^2 F)$, (d) $a_1/\lambda$, (e) $a_2/\lambda$, (f) $a_3/\lambda$, and (g) $a_6/\lambda$. Here
$2 \pi D /(\zeta \lambda F) \!=\!0.1$.
Domain of positive pinion velocity in the $(W r/(F R), \zeta \lambda \omega/(2\pi R^2 F))$ space for (h)
$2 \pi D /(\zeta \lambda F) \!=\!0$ and (i) $2 \pi D /(\zeta \lambda F) \!=\!0.2$.
Except the parameter being varied
$ 2 \pi a_1/\lambda   \!=\!  16.25 $, $ 2 \pi a_2/\lambda \!=\! 7.5 $, $ 2 \pi a_3/\lambda \!=\! 4.16 $,
$ 2 \pi a_4/\lambda \!=\! 1.875 $, $ 2 \pi a_5/\lambda \!=\! 0.5 $, $ 2 \pi a_6/\lambda \!=\! 0.83 $, and
$ \zeta \lambda \omega/(2\pi R^2 F) \!=\!0.4$ are assumed. } \label{fighexa}
\end{center}
\end{figure*}

\section{Polychromatic driving}

As an example of multi-harmonic driving signal, we consider
\begin{eqnarray}
y(t)  &\!\! \!\!=& \!\! \!\!  a_1 \! \cos (\omega t \!\!+\!    \frac{5 \pi}{3}) \!+\! a_2 \! \cos (2 \omega t \!+\!   \frac{4 \pi}{3}) \!+\!
a_3\! \cos (3 \omega t \!+\!  \pi) \nonumber \\
&\!\!\! \! + & \!\!\! \!  a_4 \cos (4\omega t \!\!+\!  \frac{2 \pi}{5}) \!+\! a_5 \cos (5\omega t \!\!+\!  \frac{\pi}{4}) \!+\!
a_6 \cos (6 \omega t \! ),  \label{tritri}
\end{eqnarray}
where  $a_1^*  \!=\! 16.25$,
$a_2^*  \!=\! 7.5$, $a_3^*  \!=\! 4.16$, $a_4^*  \!=\! 1.875$, $a_5^*  \!=\! 0.5$, $a_6^*  \!=\! 0.83$, and
$\omega^* \!=\!0.4$, see Fig.~\ref{fighexa}(a). We assume that the noise strength is $D_S \!=\!0.1$.
To study the average pinion velocity, we change {one} of the above parameters and keep others
fixed. Figures~\ref{fighexa}(b)-(g) demonstrate $V_P/V_S $ for various loads, as a function of $D_S$, $\omega^*$,
$a_1/\lambda$, $a_2/\lambda$, $a_3/\lambda$, and $a_6/\lambda$, respectively.
Figures~\ref{fighexa}(b)-(g) clearly show the possibility of choosing the parameters to guarantee $V_P>0$.
However, this domain of parameters shrinks, as the load increases. Figure~\ref{fighexa}(b) shows that for $W_S \!=\! 0.1$ and $W_S \!=\!0.15 $, indeed $V_P/V_S $ first increases, reaches its maximum, and then monotonically decreases as
the noise strength $D_S$ increases. This reminds us the stochastic resonance effect\ \cite{effect}. 
Figure~\ref{fighexa}(c) shows that $V_P/V_S$ is zero for frequencies smaller than a threshold $\omega^*_{th}$, 
cf. Fig.~\ref{figbi1}(c). We find $\omega^*_{th}= 0.018$ when $W_S=0.05$. For loads $W_S=0.05$, $ 0.1$ and $0.15$, the 
average pinion velocity becomes practically positive when $\omega^*$ becomes larger than $0.12$, $0.156$, and $0.192$, respectively. 
$V_P/V_S$ is monotonically decreasing when $\omega^*  \!>\! \omega^{*}_{\times}$, cf. Fig.~\ref{figbi1}(c).  
For example, $ \omega^{*}_{\times}= 0.33 $ when $ W_S \!=\! 0.05$, 
but shifts to $ \omega^{*}_{\times}= 0.29$ when $ W_S \!=\! 0.15$. 
The behavior of $V_P/V_S$ for $\omega^*< \omega^{*}_{\times}$ again warns that the frequency must be deliberately tuned to maximize the average pinion velocity. Figure~\ref{fighexa}(c)-(g) show that $V_P/V_S$ may be positive in {\it disconnected} windows of parameters, cf. Figs.~\ref{figbi1}(d) and~\ref{figbi1}(f). Figures~\ref{fighexa}(h) and \ref{fighexa}(i) demonstrate domains of positive pinion velocity in the $(W_S,\omega^* )$ space for
$D_S=0$ and $D_S=0.2$, respectively.
The noise may degrade the machine performance:
For example, at $\omega^*=0.04$, the maximum load decreases from $W_S=0.097 $ to $W_S=0.076$ as the temperature
increases. But the noise may facilitate the device operation: For $D_S=0$, two gaps $(0, \omega^*_{th}=0.02 )$ and $(0.06,0.12)$ indicate that the device does not lift up the load.
On increasing the temperature, the first gap shrinks to $(0, \omega^*_{th}=0.015 )$ and the second gap completely disappears, 
cf. Figs.~\ref{figbi2}(a)-(b). 

\begin{figure*}[t]\begin{center}
    \begin{minipage}[b]{0.32\textwidth}\begin{center}
        \includegraphics[width=\textwidth]{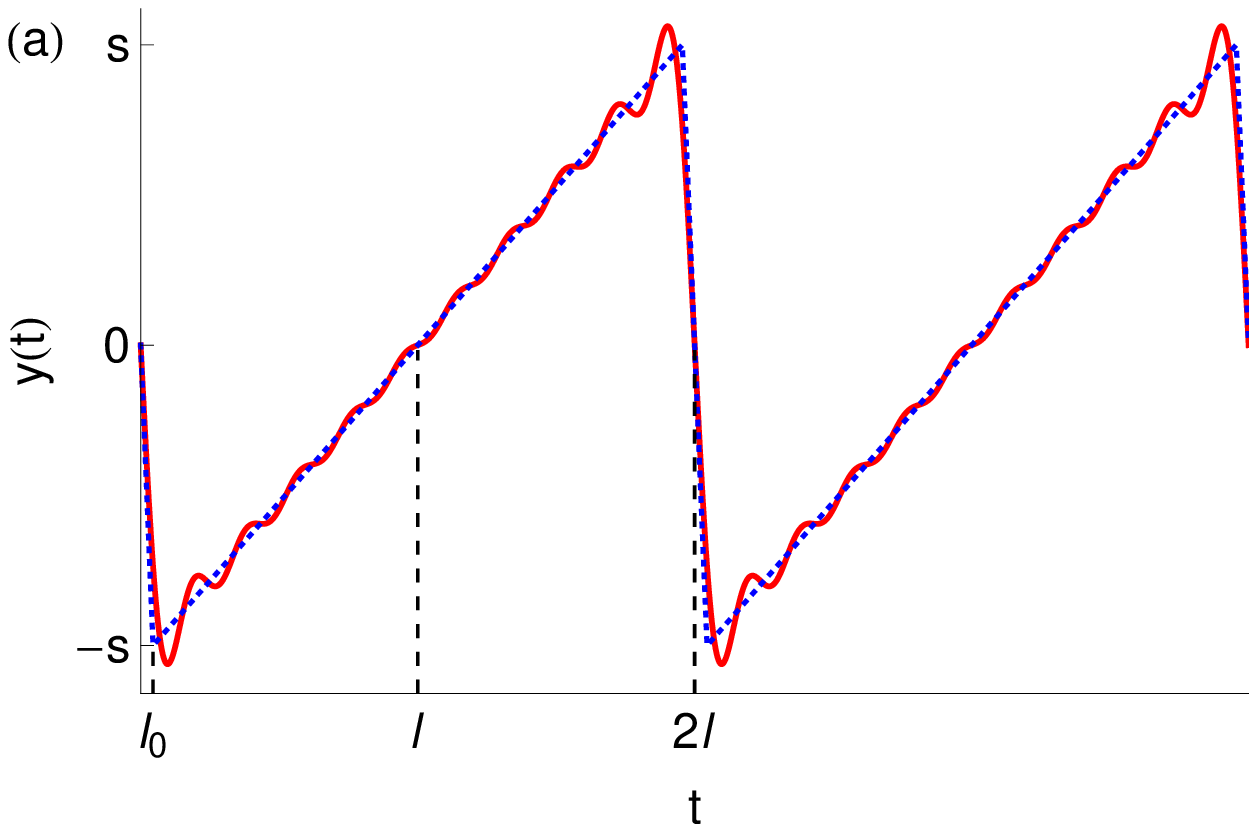}
    \end{center}\end{minipage} \hskip+0cm
    \begin{minipage}[b]{0.32\textwidth}\begin{center}
        \includegraphics[width=\textwidth]{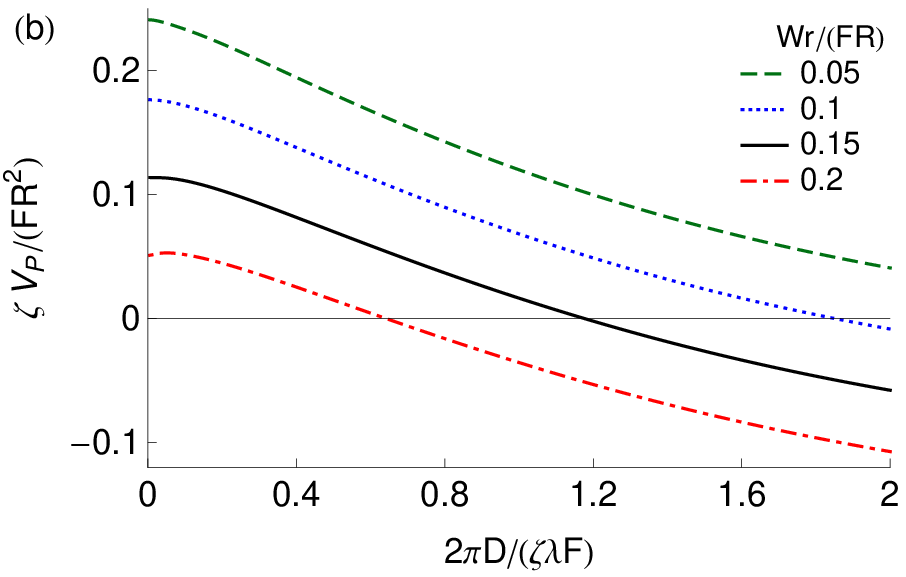}
    \end{center}\end{minipage} \hskip0cm
    \begin{minipage}[b]{0.32\textwidth}\begin{center}
        \includegraphics[width=\textwidth]{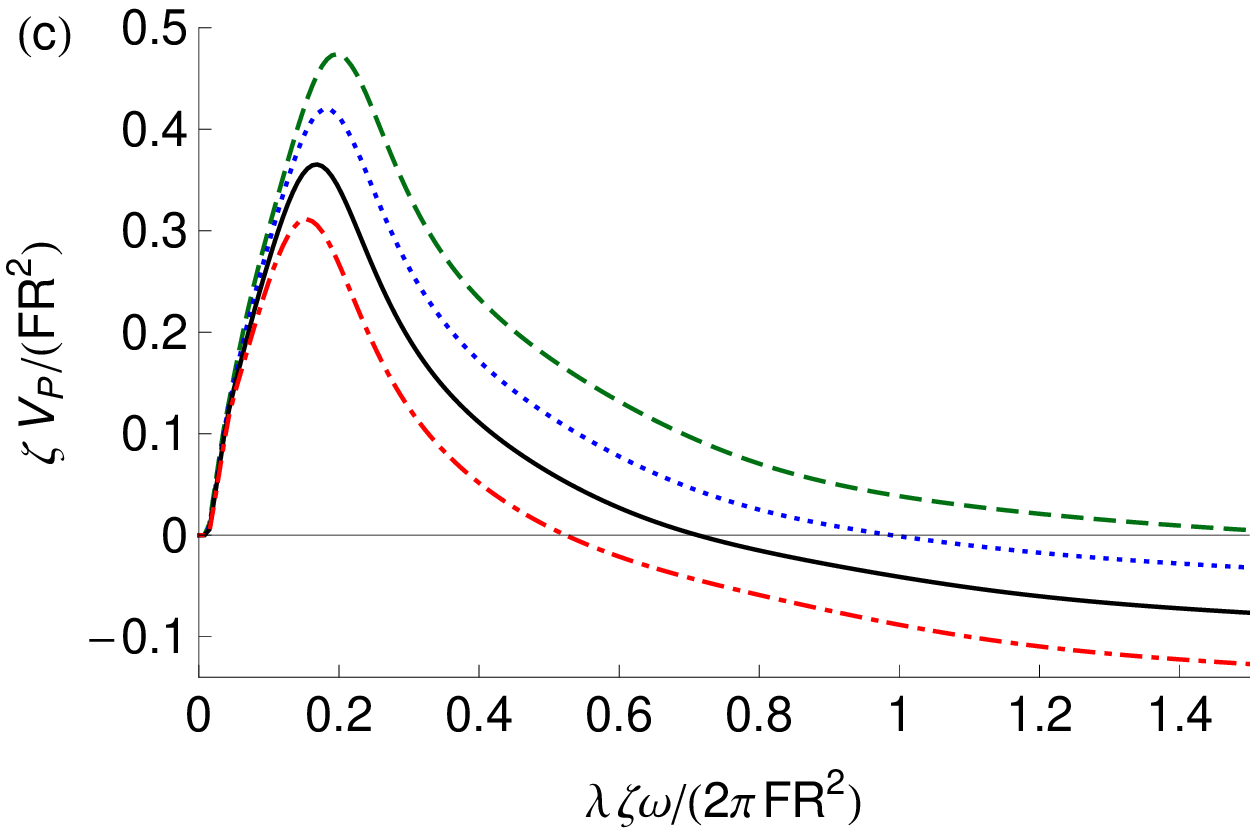}
    \end{center}\end{minipage} \hskip0cm
     \begin{minipage}[b]{0.32\textwidth}\begin{center}
        \includegraphics[width=\textwidth]{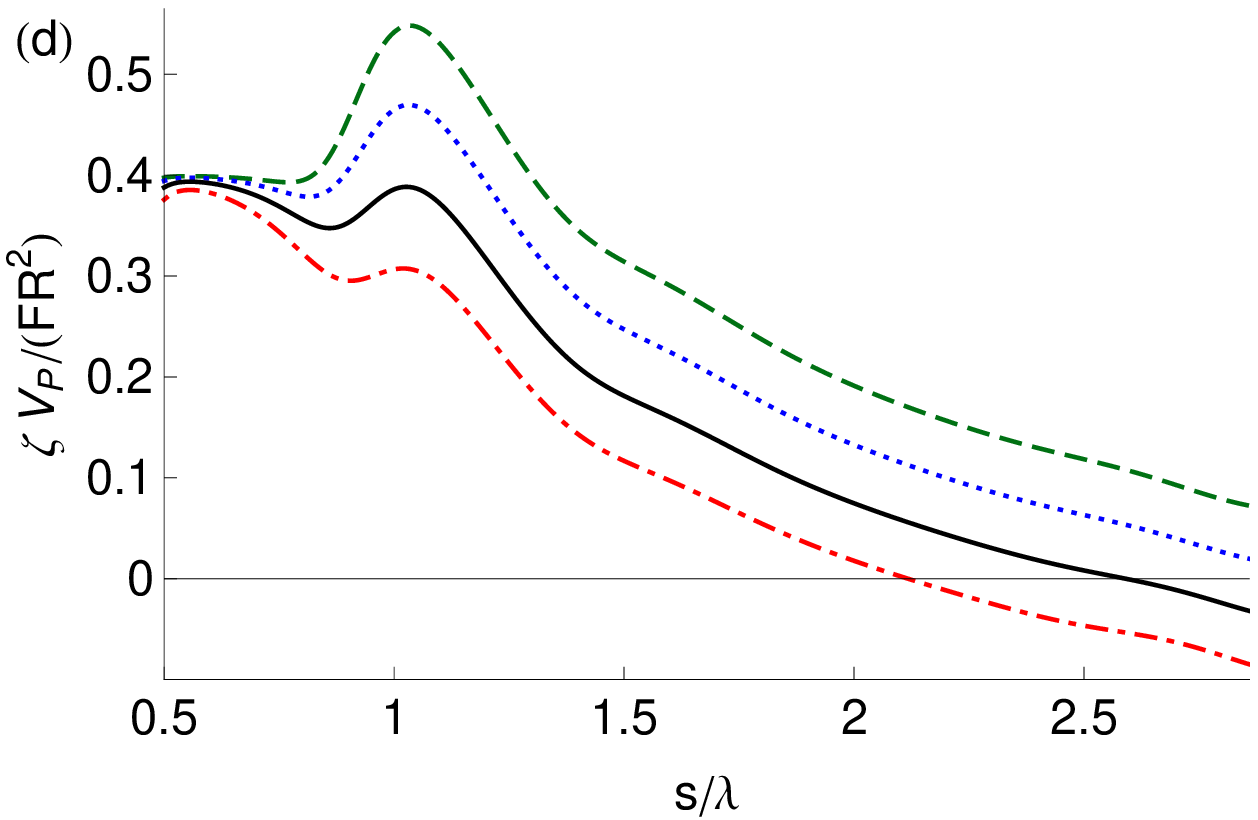}
    \end{center}\end{minipage} \hskip+0cm
    \begin{minipage}[b]{0.32\textwidth}\begin{center}
        \includegraphics[width=\textwidth]{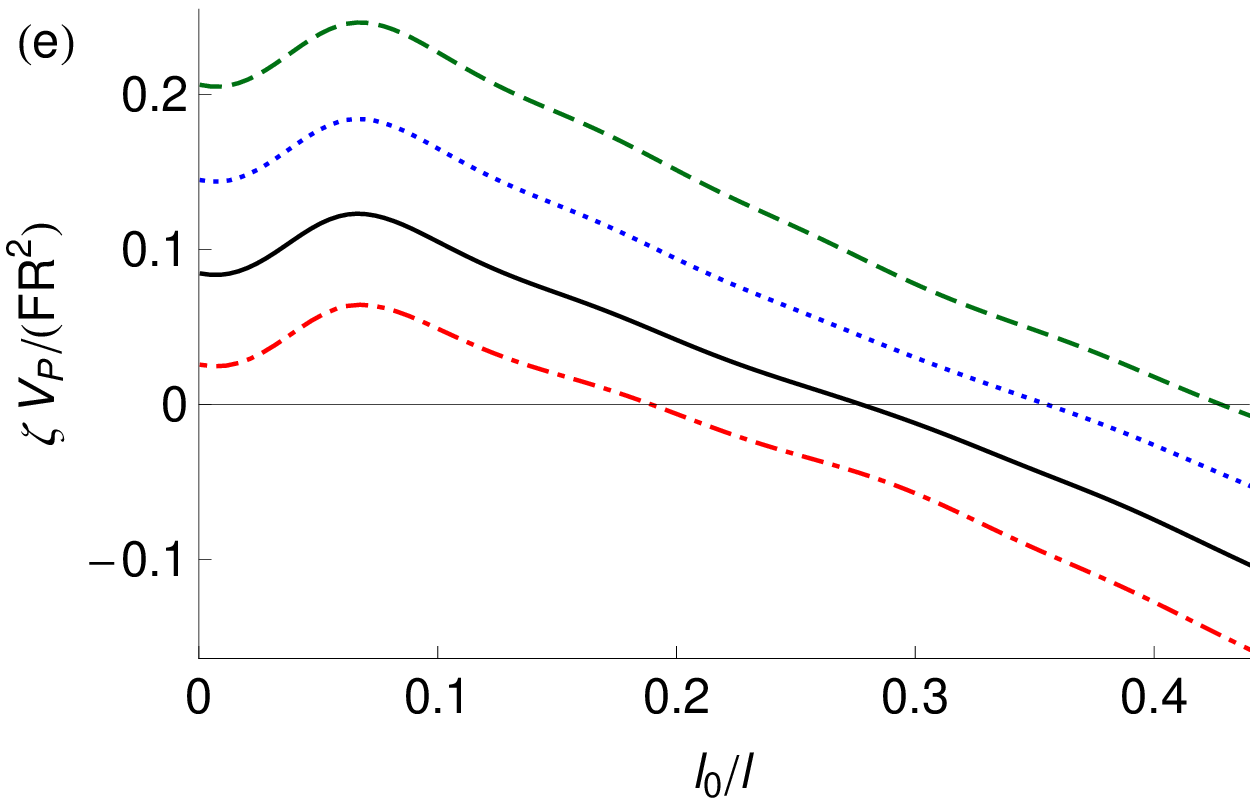}
    \end{center}\end{minipage} \hskip0cm
    \begin{minipage}[b]{0.32\textwidth}\begin{center}
        \includegraphics[width=\textwidth]{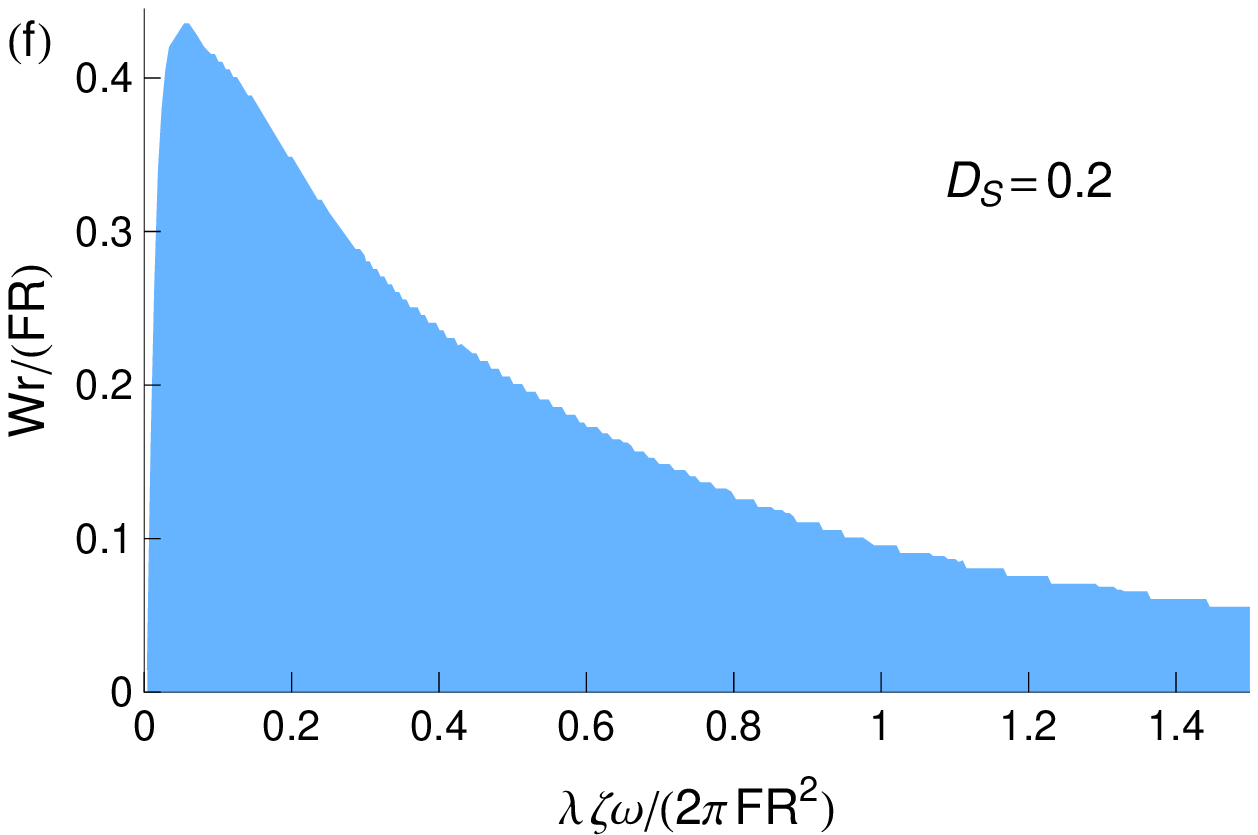}
    \end{center}\end{minipage} \hskip0cm
\caption{(Color online) (a) The periodic triangular signal
characterized by $l$, $l_0$, and $s$. $V_P$ in units of ${R^2  F}/{\zeta}$ versus (b) $2 \pi D /(\zeta \lambda F)$,
(c) $\zeta \lambda \omega/(2\pi R^2 F)$, (d) $s/\lambda$, and (e) $l_0/l$. (f) Domain of positive pinion velocity in the $(W r/(F R), \zeta \lambda \omega/(2\pi R^2 F))$ space for $ 2 \pi D /(\zeta \lambda F) \!=\!0.2$.
Except the parameter being varied, $ 2 \pi s/\lambda=11.4$,  $ \zeta \lambda \omega/(2\pi R^2 F)  \!=\!0.4$, $l_0/l \!=\!0.0446 $,
$N_h \!=\! 10$, and $ 2 \pi D /(\zeta \lambda F) \!=\!0.1$ are assumed.
 } \label{triangu}
\end{center}
\end{figure*}

As another multi-harmonic signal, we consider
\begin{equation}
y(t)= \sum_{n=1}^{N_h}   \frac{- 2 s l^2}{  \pi^2 n^2 l_0 (l-l_0)  }  \sin(\frac{n \pi l_0}{l})   \sin( \frac{n \pi t }{l}), \label{dah}
\end{equation}
which approximates the periodic triangular signal
characterized by the durations $l$ and $l_0$, and amplitude $s$, see Fig.~\ref{triangu}(a).
$\omega=\pi/l$ and $N_h$ are the fundamental frequency and the number of harmonics, respectively.
As a concrete example, we assume  $s^*=11.4$,  $\omega^* \!=\!0.4$, $l_0/l \!=\!0.0446 $, $N_h \!=\! 10$, and $D_S \!=\!0.1$.
Similar to our previous studies of $V_P$, we change one of the above parameters and keep others fixed.
Figures~\ref{triangu}(b)-(e) demonstrate $V_P/V_S $ for various loads, as a function of $D_S$, $\omega^*$, $s/\lambda$,
and $l_0/l$, respectively. We observe that the signal parameters $\omega^*$, $s/\lambda$, and even $l_0/l$ must be tuned to
maximize the average pinion velocity. Here $V_P/V_S$ can be as large as $0.5$, note that in all previous examples $V_P/V_S$ is below $0.2$. Figures~\ref{triangu}(b)-(e) again show the possibility of choosing the parameters to guarantee $V_P>0$. As expected, this domain of parameters shrinks, as the external load increases. Figure~\ref{triangu}(c) shows that $V_P/V_S$ is monotonically decreasing when $\omega^*  \!>\! \omega^{*}_{\times}$. As before,  $\omega^{*}_{\times}$ depends on the external load. Here $\omega^*_{th}=0.008$ is quite small, cf. Figs.~\ref{figbi1}(c) and~\ref{fighexa}(c). Figure~\ref{triangu}(f) demonstrates domain of positive pinion velocity in the $(W_S,\omega^* )$ space for room temperature noise $D_S=0.2$. 
Note that here loads as large as $W_S = 0.435 $ can be lifted up, while in previous examples admissible loads are below $W_S=0.2$.

\section{Remarks}\label{sec:dis}

We map the Langevin equation (\ref{asli}) describing the pinion dynamics into the Eq.~(\ref{moaser}) which is well known in the realm of Brownian motors\  \cite{rev2,rev}. However, one must note that
(i) The Casimir device is intended to rectify a {\it deterministic} periodic input motion (rack motion).
(ii) Only {\it positive} average pinion velocities are of interest, since the device should lift up the load.
This restricts the allowed set of parameters $ (W_S,D_S,\omega^*,a_1^*, \phi_1, a_2^*,\phi_2,..., a_{N_h}^*, \phi_{N_h}) $  characterizing the system. One can use the stochastic Runge-Kutta algorithm of the second order\ \cite{honeycutt}
or similar methods to solve the Langevin equation (\ref{moaser}). This is straightforward, but requires a large ensemble of realizations to obtain a reliable average. We follow Denisov, H\"{a}nggi and Mateos~\cite{Denisov} to map the original problem onto a set of linear algebraic equations. This allows us to explore a huge parameter space in a reasonable computational time (a few days).

Input signals that lead to $V_P \leqslant 0$ are of no interest. It is known that $V_P=0$\ \cite{ref9,ref22,ref23} if
$g(z^*, t^*) \rightarrow - g(z^*, t^*)$ under one of the transformations
\begin{eqnarray}
T_1 : & &    z^* \rightarrow - z^*  + z^*_0    ~~~~ t^* \rightarrow + t^* + t^*_0 , \\ \nonumber
T_2 : & &    z^* \rightarrow  +z^*  + z^*_0    ~~~~ t^* \rightarrow - t^* + t^*_0,
\end{eqnarray}
specified by the parameters $z^*_0$ and $t^*_0$. According to Eq.~(\ref{myg}),
both symmetries $T_1$ and $T_2$ are absent if $W_S \neq 0$.
For $W_S \!=\! 0$ and $y(t)= a_1 \sin (\omega t +\varphi_1) $, the symmetry $T_1$ exists and thus $V_P=0$.
Now it is not surprising that the device can not lift up a load upon monochromatic driving of the rack.
For $W_S = 0$ and biharmonic driving (\ref{twotwo}), the symmetry $T_2$ exists if $\phi= \pi/4,3\pi/4,5\pi/4,...$. Figure~\ref{figbi1}(f) shows that at $\phi= \pi/4,3\pi/4,5\pi/4,7\pi/4 $ indeed
$V_P \rightarrow 0$ as $W_S \rightarrow 0$. In the case of multiharmonic signals (\ref{tritri}) and (\ref{dah}), both symmetries $T_1$ and $T_2$ are absent even if $W_S = 0$. For similar studies of Brownian motors driven by biharmonic signals, see Refs.\ \cite{Denisov,bi1,bi2,bi3,bi4}.

\begin{figure}
\includegraphics[width=0.7\columnwidth]{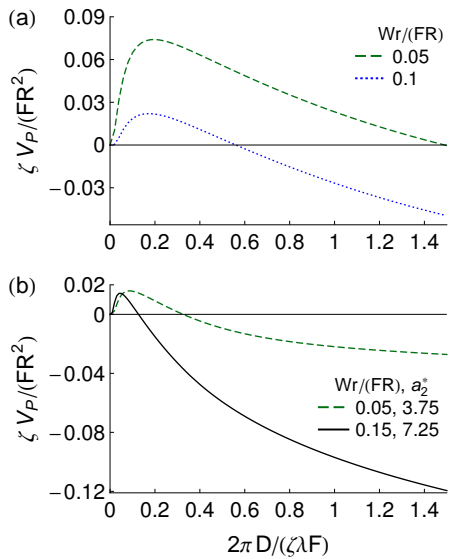}
\caption{(Color online) $V_P$ in units of ${R^2  F}/{\zeta}$ versus $2 \pi D /(\zeta \lambda F)$ for 
(a) the biharmonic driving~(\ref{twotwo}) with parameters 
$\omega^* \!=\!0.2$, $a_1^* \!=\!13.5$, $a_2^* \!=\!4.075$, and $\phi \!=\!\pi/2$, 
(b) the polyharmonic signal~(\ref{tritri}) with parameters 
$ \omega^* \!=\!0.4$, $a_1^* \!=\!16.25$, $a_3^*\!=\!4.16$, $a_4^* \!=\!1.875$, $a_5^*\!=\!0.5$, and $a_6^*\!=\!0.86$. 
} \label{stores}
\end{figure}

We observe that the noise may facilitate the device operation.
Figures~\ref{figbi2}(a)-(f) and~\ref{fighexa}(h)-(i) show that the gaps in the $V_P\!>\!0$ domain of parameters, close as the
temperature increases. This suggests to search for the fingerprints of the stochastic resonance effect in the
gaps of $V_P\!>\!0$ domain of parameters. For the biharmonic driving,
$(W_S \!=\!0.05, 0.1,\omega^* \!=\!0.2,a_1^* \!=\!13.5, a_2^* \!=\!4.075 ,\phi \!=\!\pi/2)$ are in the gap, see Fig~\ref{figbi2}(c).
Figure~\ref{stores}(a) clearly supports the occurrence of stochastic resonance for these parameters.
Figure~\ref{stores}(b) shows the occurrence of stochastic resonance for the polyharmonic signal~(\ref{tritri}). 
Here $ (\omega^* \!=\!0.4, a_1^* \!=\!16.25, a_3^*\!=\!4.16, a_4^* \!=\!1.875, a_5^*\!=\!0.5, a_6^*\!=\!0.86)$, and 
dashed and solid lines correspond to $(W_S \!=\!0.05, a_2^* \!=\! 3.75)$ and $(W_S \!=\!0.15, a_2^* \!=\! 7.25)$, 
respectively. Note that $V_P\!=\!0$ for $D_S \!= \!0$.

Our results show that  $V_P/V_S \! \sim\! 0.1-0.2 $ is achievable, see Figs.~\ref{figbi1},~\ref{fighexa} and~\ref{triangu}.
Reminding the typical pinion radius $R \!=\! 1 \;\mu{\rm m}$ and skipping velocity
$ V_S  \!=\!{R^2  F}/{\zeta} \!=\!  3.75  \;\mu{\rm m}/{\rm s} $ for the gap $H=200 \;{\rm nm}$, this means that the pinion rotates with an average angular velocity $V_P/R \sim 0.4- 0.8\;{\rm Hz} $.
Quite importantly, the amplitude of the lateral Casimir force rapidly grows as the gap $H$ decreases.
For $H=100 \;{\rm nm}$ then $ V_S = 146.25   \;\mu{\rm m}/{\rm s} $ and $V_P/R \sim 16- 32 \;{\rm Hz} $.
This is a considerable average angular velocity.

According to Eq.~(\ref{myds}), the effective noise strength $D_S \propto 1/F$ rapidly decreases as the gap size $H$ decreases.
Assuming $H=200 \;{\rm nm}$, we estimated $D_S=0.18$ at room temperature.
Now reducing the gap size by only a factor of two yields $D_S= 0.005$.
Thus the noise influence on the the device operation can be controlled and optimized.

In summary, the Casimir rack and pinion acts as a nano-scale mechanical rectifier:
Even in the presence of the thermal noise, the device can lift up a load when the rack undergoes a periodic multiharmonic motion.
The geometrical parameters, especially the gap $H$, and the input signal parameters, can be tuned to optimize the device performance.

In light of recent measurements of lateral Casimir force~\cite{Mohideen,c2010}, experimental
realization of the Casimir rack and pinion is not out of reach.

\begin{acknowledgments}
We thank financial support from the
“Center of Excellence on the Structure and Physical
Properties of Matter” of the University of Tehran.
We also thank R. Sepehrinia and A. A. Saberi for their enlightening comments.
\end{acknowledgments}

\end{document}